\documentclass{aa}

\usepackage{graphicx}
\usepackage[varg]{txfonts}
\newcommand{\angstrom}{\textup{\AA}}

\newcommand\T{\rule{0pt}{2.6ex}}       
\newcommand\B{\rule[-1.2ex]{0pt}{0pt}} 

\begin{document}

\title{The halo of M105 and its group environment as traced by planetary
nebula populations\thanks{Based on data collected at Subaru Telescope, which is operated by the
National Astronomical Observatory of Japan under programme S14A-006 and with
the William Herschel Telescope operated on the island of La Palma by the
Isaac Newton Group of Telescopes in the Spanish Observatorio del Roque de los
Muchachos of the Instituto de Astrof\'{i}sica de Canarias.}}

\subtitle{I. Wide-field photometric survey of planetary nebulae in the Leo~I group}
\author{J. Hartke\inst{1,2} \and
       M. Arnaboldi\inst{1} \and
       O. Gerhard\inst{3} \and
       L. Coccato\inst{1} \and
       C. Pulsoni\inst{3,4} \and
       K. C. Freeman\inst{5} \and
       M. Merrifield\inst{6} \and
       A. Cortesi\inst{7} \and
       K. Kuijken\inst{8}
       }

\institute{European Southern Observatory,
       Karl-Schwarzschild-Str. 2, 85748 Garching, Germany
       \and
       European Southern Observatory,
       Alonso de Cordova 3107, Vitacura,
       Casilla 19001, Santiago de Chile, Chile\\
       \email{jhartke@eso.org}
       \and
       Max-Planck-Institut f\"{u}r Extraterrestrische Physik,
       Giessenbachstra{\ss}e, 85748 Garching, Germany
       \and
       Excellence Cluster Universe,
       Boltzmannstra{\ss}e 2, 85748, Garching, Germany
       \and
       Research School of Astronomy \& Astrophysics Mount Stromlo
       Observatory, Cotter Road,
       2611 Canberra, Australia
       \and
            School of Physics and Astronomy,
       University of Nottingham, NG7 2RD, UK
       \and
       Departamento de Astronomia,
       Instituto de Astronomia, Geofisica e Ciencias Atmosfericas da USP,
       Cidade Universitaria,
       CEP:05508900 Sao Paulo, Brazil
       \and
       Leiden Observatory, Leiden University,
       PO Box 9513, 2300~RA Leiden, The Netherlands
       }

\date{\today}

\abstract{M105 (NGC~3379) is an early-type galaxy in the Leo~I group. The Leo~I group is the nearest group that contains all main galaxy types and can thus be used as a benchmark to study the properties of the intra-group light (IGL) in low-mass groups.}
{We present a photometric survey of planetary nebulae (PNe) in the extended halo of the galaxy to characterise its PN populations and investigate  the presence of an extended PN population associated with the intra-group light.}
{We use PNe as discrete stellar tracers of the diffuse light around M105. These PNe were identified on the basis of their bright [\ion{O}{iii}]5007\AA\ emission and the absence of a broad-band continuum using automated detection techniques. We compare the PN number density profile with the galaxy surface-brightness profile decomposed into metallicity components using published photometry of the Hubble Space Telescope in two halo fields.}
{We identify 226 PNe candidates within a limiting magnitude of $m_{5007, \mathrm{lim}} = 28.1$ from our Subaru-SuprimeCam imaging, covering 67.6~kpc (23 effective radii) along the major axis of M105 and the halos of NGC~3384 and NGC~3398. We find an excess of PNe at large radii compared to the stellar surface brightness profile from broad-band surveys. This excess is related to a variation in the luminosity-specific PN number $\alpha$ with radius. The $\alpha$-parameter value of the extended halo is more than seven times higher than that of the inner halo. We also measure an increase  in the slope of the PN luminosity function at fainter magnitudes with radius.}
{We infer that the radial variation of the PN population properties is due to a diffuse population of metal-poor stars ([M/H] $< -1.0$) following an exponential profile, in addition to the M105 halo. The spatial coincidence between the number density profile of these metal-poor stars and the increase in the $\alpha$-parameter value with radius establishes the missing link between metallicity and the post-asymptotic giant branch phases of stellar evolution. We estimate that the total bolometric luminosity associated with the exponential IGL population is $2.04 \times 10^{9} \; L_{\odot}$ as a lower limit. The lower limit on the IGL fraction is thus 3.8\%. This work sets the stage for kinematic studies of the IGL in low-mass groups.}
\maketitle
%

\section{Introduction}

In a cold dark matter (CDM) dominated Universe with a cosmological constant ($\Lambda$CDM), structure forms hierarchically \citep{1978MNRAS.183..341W,1980lssu.book.....P}. In this paradigm, the favoured formation scenario for early-type galaxies (ETGs) is a two-phase process: the initial phase of strong in situ star formation \citep[e.g.][and references therein]{2005MNRAS.360.1355T} is followed by growth through mergers and accretion \citep{1991ApJ...379...52W,2002NewA....7..155S,2010ApJ...725.2312O}. Therefore, the study of the distribution and kinematics of stars in their extended halos can provide important constraints on late galaxy growth, as accretion events leave behind long-lasting dynamical signatures in the velocity phase-space \citep[e.g.][]{2005ApJ...635..931B, 2008AJ....135.1998R, 2009ApJ...698..567S, 2011ApJ...733L...7H, 2016ApJ...823...19C} as well as low-surface brightness features such as stellar streams, fans, and plumes \citep[e.g.][]{2002AJ....124.1452F, 2005ApJ...631L..41M, 2007ApJ...658..337B, 2010ApJ...715..972J, 2017ApJ...834...16M, 2015A&A...579L...3L}.

The galaxy M105 (NGC~3379) has long been regarded as a prototypical ETG to be described by the classic de Vaucouleurs surface brightness profile \citep{1979ApJS...40..699D}.
Later, M105 was at the centre of a long-standing debate on the dark matter (DM) content of ETGs and is one of the poster children of the Planetary Nebula Spectrograph (PN.S) ETG survey \citep{2007ApJ...664..257D}. Using data obtained with the PN.S, \citet{2003Sci...301.1696R} and \citet{2007ApJ...664..257D} found that several intermediate-luminosity elliptical galaxies, one of which is M105, appeared to have low-mass and low-concentration DM halos, if any, on the basis of their rapidly falling velocity dispersion profiles.
Follow-up studies argued that this apparent lack of dark matter might be due to anisotropy and viewing-angle effects \citep{2003Sci...301.1696R, 2005Natur.437..707D, 2009MNRAS.395...76D}. Similarly, Schwarzschild modelling based on \textsc{Sauron} integral-field spectroscopic data was also consistent with a DM halo and non-negligible radial anisotropy \citep{2009mnras.398..561w}.

M105 is not an isolated galaxy but is situated in a group environment. The Leo~I group, named initially G11 group, in which M105 resides, is the nearest loose group to contain both early- and late-type massive galaxies \citep{1975gaun.book..557D}. The group contains at least 11 bright galaxies and has an on-sky extent of $1.6\times1.0\;\mathrm{Mpc}$ \citep{1975gaun.book..557D}. Of the 11 bright galaxies, 7, including M105, are associated with the M96 (NGC~3368) group, and the remaining 4 are associated with the so-called Leo Triplet.
The Leo~I group is rich in dwarf galaxies: \citet{2018A&A...615A.105M} recently discovered 36 new candidates in a 500 square-degree volume in addition to the previously known 52 dwarf galaxies.

The group is surrounded by a segmented ring with a diameter of 200~kpc that consists of neutral hydrogen \citep{1983ApJ...273L...1S}. After its discovery, \citet{1985ApJ...288L..33S} showed that the kinematics of the \ion{H}{i} gas could be reconciled with an elliptical orbit centred at the luminosity-weighted centroid of M105 and the nearby NGC~3384 spiral galaxy. In their model, the ring rotates with a period of approximately 4~Gyr.
However, the crossing times in the group are much shorter than this \citep{1985AJ.....90..450P}, which is problematic for the stability of the ring. Moreover, only the disc of NGC~3384 is aligned with the orientation of the ring, which is puzzling if M105, NGC~3384, and M96 formed from a single rotating cloud.
An alternative formation scenario to the primordial origin of the \ion{H}{i} ring is the collision of two spiral galaxies \citep{1984ApJ...285L...5R,1985ApJ...288..535R}. A recent simulation of the collision of two gaseous disc galaxies reproduces both the shape and rotation of the \ion{H}{i} ring and the lack of a visible-light counterpart \citep{2010ApJ...717L.143M}.

There is no evidence for low surface brightness (LSB) features associated with the \ion{H}{i} ring and for an extended diffuse intra-group light (IGL) out to the HI ring, brighter than the surface brightness (SB) limit of $\mu_B = 30\;\mathrm{mag}\;\mathrm{arcsec}$ \citep{2014ApJ...791...38W}.
Furthermore, \citet{2003A&A...405..803C} did not detect any PNe associated with the \ion{H}{i} ring and set an upper limit of $1.6^{+3.4}_{-1.0}\%$ on the IGL fraction, that is, on the fraction of light attributed to the IGL with respect to the light from group galaxies. These findings contradict the results from numerical simulations, which predict that galaxy groups have IGL fractions between $12\%$ and $45\%$ \citep{2006MNRAS.369..958S,2006ApJ...648..936R}.
However, \citet{2014ApJ...791...38W} argued that galaxies in the Leo~I group fall below the mass resolution of the simulations mentioned above, and we might thus expect a correspondingly dimmer IGL component. If this were the case, \citet{2014ApJ...791...38W} argued that the IGL would still contribute a few percent at most towards the total light in the group.

\citet{2007ApJ...666..903H} found evidence for a metal-poor (MP) halo at $\sim 12\;R_\mathrm{eff}$ based on deep and high spatial resolution Hubble Space Telescope (HST) photometry in the western halo of M105. Their Advanced Camera for Surveys (ACS) field seems to lie at the transition radius where the MP red giant branch (RGB) population starts to dominate the metal-rich (MR) one. Both populations are inferred to have ages of 12~Gyr and metallicities of $\mathrm{[M/H]} < -0.7$ and $\mathrm{[M/H]} \geq -0.7,$ respectively.
\citet{2016ApJ...822...70L} combined these data with an archival HST ACS pointing in the very inner halo at approximately $3~R_\mathrm{eff}$ south-east from the centre of M105. In their independent analysis, they corroborated the existence of two distinct subpopulations both in colour and metallicity: a dominant, red, MR population with an approximately solar peak metallicity, and a much weaker, blue, MP population whose peak metallicity is $[\mathrm{M/H}] \approx - 1.1$.
The average ages and metallicities in the inner halo agree with those inferred by \citet{2009mnras.398..561w}, who found a stellar population consistent with a metallicity slightly below 20\%\  solar metallicity ($\approx -0.7$ dex) and an age of 12 Gyr.

The origin of this extended MP halo is still a subject of debate. \citet{2007ApJ...666..903H} ruled out a \emph{\textup{single}} chemical evolution sequence for the halo stars. Instead, they proposed a multi-stage formation model: the MP halo was built up first from pristine gas, and the star formation therein was truncated approximately at the time of cosmological reionisation. The MR, bulge-like component was formed later from pre-enriched gas and with a higher yield, eventually resulting in the observed solar abundances.
\citet{2016ApJ...822...70L} also suggested a two-phase formation model for the halo in which the red and MR halo was formed in situ or through the major mergers of massive progenitors, while the blue and MP halo was formed through dissipationless mergers and accretion.

Using extended photometric and kinematic samples of planetary nebulae (PNe), we aim to place better constraints on the assembly history of M105 and its group environment. Our new PN data span a minor-axis distance that is three times larger than that of \citet{2007ApJ...664..257D}. This will place valuable new constraints on the stellar populations in the outer halo as traced by PNe, and on the PN spatial and line-of-sight (LOS) velocity distributions of the M105 stars at large radii.

This paper is the first in a series and is organised as follows. In Sect.~\ref{sec:dataM105} we present the photometric and kinematic surveys of PNe in the Leo~I group. In Sect.~\ref{sec:M105SB} we review the results from broad-band photometric surveys. We then study the relation between the azimuthally averaged distribution of the PNe and the SB photometry in Sect.~\ref{sec:105_alpha}. In Section~\ref{sect:RGB} we evaluate the constraints from the resolved stellar populations and relate their physical parameters to the observed properties of the PN sample in M105.  Section~\ref{sec:PNLFM105} covers the PN luminosity function (PNLF) of M105 and its variation with radius. We conclude this paper with a discussion in Sect.~\ref{sec:M105disc}. Paper~II in this series will focus on the dynamics of the M105 halo and the Leo~I group.

In the remainder of this paper, we adopt a physical distance of 10.3~Mpc to M105 from surface brightness fluctuation measurements \citep[SBF;][]{2001ApJ...546..681T}, which agrees well with the tip of the red giant branch (TRGB) distance independently determined by \citet{2007AJ....134...43H,2007ApJ...666..903H} and \citet{2016ApJ...822...70L}.
The corresponding physical scale is $49.6\;\mathrm{pc}/\arcsec$. The effective radius of M105 has been determined to be $r_{\mathrm{e,M105}} = 54\farcs8 \pm 3\farcs5$, which corresponds to a physical radius of 2.7~kpc \citep{1990AJ.....99.1813C}. The structural parameters of M105 relevant to this paper are summarised in Table~\ref{tabM105:galproperties}.

\section{Photometric and kinematic surveys of PNe in the Leo~I group}
\label{sec:dataM105}
\begin{figure*}
  \centering
    \includegraphics[width=18cm]{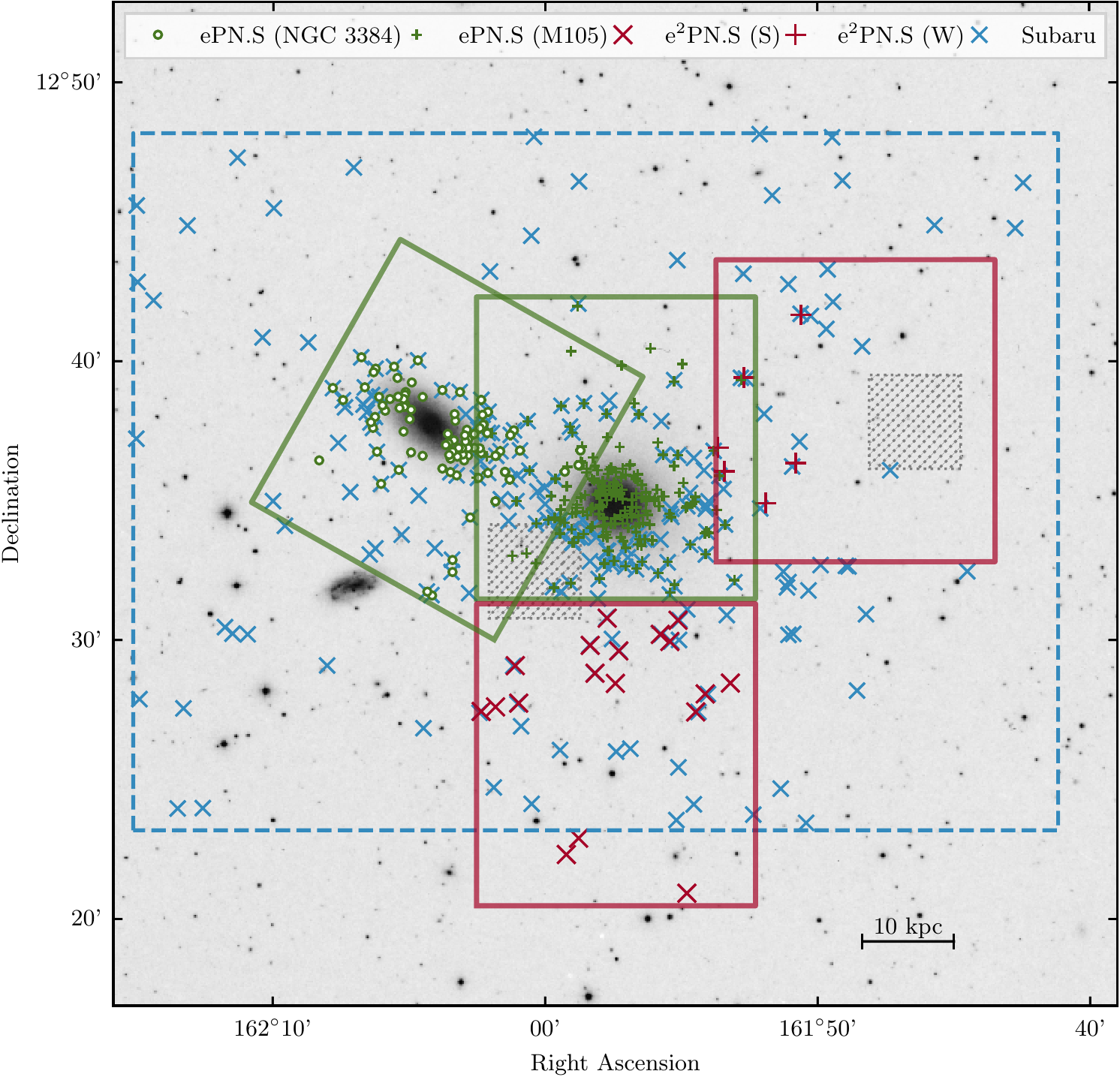}
  \caption{DSS image of M105. Overplotted are the PN candidates from the Subaru survey (blue crosses), PNe from the ePN.S survey \citep[][green pluses]{2018A&A...618A..94P}, and the e$^{2}$PN.S survey (this work, red pluses and crosses). PNe in NGC~3384 from \citep{2013A&A...549A.115C} are denoted with green circles. The red and green rectangles denote the ePN.S and e$^{2}$PN.S fields, and the dashed blue rectangle indicates the SuprimeCam footprint. The grey hatched regions indicate the two HST fields analysed in   \citet{2016ApJ...822...70L}. The scale bar in the lower right corner denotes   10~kpc. North is up, and east is to the left.}
  \label{figM105:survey}
\end{figure*}

The photometric and kinematic surveys in this work expand our ongoing effort to study PNe as tracers in group and cluster environments. We combine kinematic data from the extended PN.S (ePN.S) ETG survey \citep{2017IAUS..323..279A,2018A&A...618A..94P}, including two additional newly acquired fields, with accurate narrow- and broad-band photometry obtained with SuprimeCam mounted at the prime focus of the $8.2\,\mathrm{m}$ Subaru telescope \citep{2002PASJ...54..833M}. In what follows, we focus on the results from our SuprimeCam survey and use the ePN.S data set as an independent validation of the PN candidates from the SuprimeCam data. A detailed analysis of the more extended kinematic PN survey will be discussed in Paper~II of this series.

\subsection{Subaru SuprimeCam photometry}
\subsubsection{Survey description and data reduction}
\label{ssec:photsurvey}
We used the on-off band technique to detect PNe candidates in M105. The pointing centred on M105 was observed through a narrow-band [\ion{O}{iii}] filter ($\lambda_\mathrm{c,on} = 5500\,\angstrom, \; \Delta \lambda_\mathrm{on} = 74\,\angstrom$) and a broad-band $V$-filter ($\lambda_\mathrm{c,off} = 5029\,\angstrom, \; \Delta \lambda_\mathrm{off} = 956\,\angstrom$), using the same observational strategy as \citet{2013A&A...558A..42L}, who observed M87 in the Virgo cluster. The observations were taken in the same night as those of \citeauthor{2017A&A...603A.104H} \citeyearpar{2017A&A...603A.104H} in their survey of M49. We therefore only briefly recapitulate the data reduction steps and refer to \citet{2017A&A...603A.104H} for further details.

In order to reach fluxes in [OIII] that are two magnitudes fainter than the PNLF bright cut-off \citep[$m^{\star}_{5007} = 25.5$, assuming an absolute bright cut-off of the PNLF of $M^{\star} = -4.51$, ][]{2002apj...577...31c}, six dithered exposures with a total exposure time of $1.8\,\mathrm{h}$ through the on-band and six dithered exposures with a total exposure time of $0.6\,\mathrm{h}$ through the off-band filter were observed. The observing conditions were photometric, with seeing better than $0\farcs7$. The photometric zero-points were calculated to be $Z_{\mathrm{[\ion{O}{iii}]},AB} = 24.51 \pm 0.04$ and $Z_{V,AB} = 27.48 \pm 0.01$ by \citet{2017A&A...603A.104H}.
To convert $AB$ magnitudes into $m_{5007}$ magnitudes from the [\ion{O}{iii}] line, we used the relation derived by \citet{2003AJ....125..514A} for the SuprimeCam narrow-band filter,
\begin{equation}
  m_{5007} = m_{AB} + 2.49.
\end{equation}
The relation between integrated flux $F_{5007}$ and magnitude $m_{5007}$ is given by the Jacoby relation \citep{1989ApJ...339...39J},
\begin{equation}
  m_{5007} = -2.5\log_{10}(F_{5007}) - 13.74.
\end{equation}

The data were reduced using the instrument pipeline SDFRED2\footnote{\url{https://www.naoj.org/Observing/Instruments/SCam/sdfred/sdfred2.html.en}}.
As the field is dominated by the three bright and extended galaxies M105, NGC~3384, and NGC~3398, the final images were astrometrised and stacked with the tools Weight Watcher \citep{2008ASPC..394..619M}, \textsc{Scamp} \citep{2006ASPC..351..112B}, and \textsc{Swarp} \citep{2002ASPC..281..228B} from the astromatic software suite. We used the UCAC-4 catalogue \citep{2012yCat.1322....0Z} as an astrometric reference.
To determine the point-spread functions (PSFs) of the images, we used the \texttt{psf} task of the \textsc{Iraf}-\texttt{daophot}\footnote{\textsc{iraf} is distributed by the National Optical Astronomy Observatories, which are operated by the Association of Universities for Research in Astronomy, Inc., under cooperative agreement with the National Science Foundation.} package.
The best-fit PSF profile is a Moffat function with $\tilde{\alpha} = 2.5$, seeing radius $r_\mathrm{FWHM} = 2.35\,\mathrm{pix}$ (corresponding to $0\farcs47$), and axial ratio $b/a = 0.96$.

\subsubsection{Selection of PN candidates and catalogue extraction}
To select PN candidates, we made use of the characteristic spectral features of PNe: a bright [\ion{O}{iii}]$5007\angstrom$ emission line and no significant continuum emission. Observed through the on-band filter, PNe appear as unresolved sources at extragalactic distances, while they are are not detected in the off-band image. \citet{2002AJ....123..760A, 2003AJ....125..514A} developed a CMD-based automatic selection procedure that selects PNe based on their excess in $\mathrm{[\ion{O}{iii}]} - V$ colour and their point-like appearance. This technique has since been tailored to SuprimeCam data by \citet{2013A&A...558A..42L} and \citet{2017A&A...603A.104H}.
For reference, the detailed catalogue extraction procedure for M105 can be found in Appendix~\ref{app:catalog_extraction}.

The limiting magnitude of our survey is defined as the magnitude at which the recovery fraction of a simulated PN population extracted from our images falls below 50\% (see Appendix \ref{app:limmag}).
We extracted 226 objects within a limiting magnitude of $m_{5007, \mathrm{lim}} = 28.1$ and therefore cover 2.6~magnitudes from the bright cut-off of the PNLF $m^{\star} = 25.5$~mag. We further discuss the nature of the PNLF of M105 in Sect.~\ref{sec:PNLFM105}. Our survey covers 67.6~kpc ($23$ effective radii) along the major axis of M105 and also covers the halos of NGC~3384 and NGC~3398. The selected PN candidates are denoted as blue crosses in Fig.~\ref{figM105:survey}.

\subsection{Extended Planetary Nebula Spectrograph (ePN.S) ETG survey}

The PN.S is a custom-built double-arm slitless spectrograph mounted at the William Herschel Telescope \citep{2002PASP..114.1234D}. Its design enables the identification of PNe and the measurement of their LOS velocities in a single observation. The ePN.S ETG survey \citep{2017IAUS..323..279A, 2018A&A...618A..94P} provides positions (RA, dec), LOS velocities ($v$), and magnitudes ($m_\mathrm{5007,PN.S}$) for spatially extended samples of PNe in 33 ETGs. The survey typically extends out to 6 effective radii ($r_\mathrm{e}$).

\subsubsection{M105 (NGC~3379)}
M105 was one of the first ETGs observed with the PN.S and was a milestone for the development of the PN.S data reduction pipeline \citep{2007ApJ...664..257D}. As part of the PN.S ETG survey \citep{2007ApJ...664..257D}, $214$ PNe were observed, covering a major-axis distance of $5.3~R_\mathrm{eff}$. These are indicated by green pluses in Fig.~\ref{figM105:survey}.

\subsubsection{NGC~3384}

As the on-sky distance between M105 and the SB0-galaxy NGC~3384 is only $435\arcsec$ and the difference in distance modulus between the galaxies is only $~0.2\;\mathrm{mag}$, their PN populations appear superimposed. An immediate separation in velocity space is not possible due to the small difference in systemic velocity ($\Delta v_\mathrm{sys} < 200\;\mathrm{km}\;\mathrm{s}^{-1}$). At the time of publication of the data \citeyearpar{2007ApJ...664..257D}, \citeauthor{2007ApJ...664..257D} addressed this by dividing the sample into two components on a probabilistic basis assuming that the distributions of PNe in the two galaxies directly follow the diffuse starlight.

The PN.S survey of S0 galaxy kinematics \citep{2013A&A...549A.115C}, also a subsample of the ePN.S ETG survey, contains $101$ PNe observed within a major-axis radius of $6.8~R_\mathrm{eff}$ from the centre of NGC~3384. Open green circles denote these in Fig.~\ref{figM105:survey}. The availability of these data will enable an improved photo-kinematic decomposition of the PN.S sample that will be described in Paper~II of this series.

\subsection{Extremely extended Planetary Nebula Spectrograph (e$^{2}$PN.S)
ETG survey in M105 and NGC 3384}

In addition to the data from the ePN.S survey, we acquired two new fields with the PN.S in March 2017 that are located west and south of M105. These two new fields, together with the two ePN.S fields centred on M105 and NGC~3384, are part of the new \textup{extremely} extended e$^{2}$PN.S ETG survey, named for its unprecedented depth and spatial coverage. The western field was positioned such that it encompassed the HST field observed by \citet{2016ApJ...822...70L}. The southern field was positioned such that the coverage along the minor axis of M105 was maximised. We observed the western field for 4.5 hours (in nine dithered exposures) and the southern field for 3.5 hours (in seven dithered exposures). The data reduction entailing PNe detection and LOS velocity and magnitude measurements was carried out using the procedures described in \citet{2007ApJ...664..257D}. The 21 newly detected PNe are denoted as red crosses in Fig.~\ref{figM105:survey}. In addition to this, we observed 3 PNe that are also part of the ePN.S survey and provide calibration for any velocity zero-point offset. The e$^{2}$PN.S survey reaches down to $1.5\;\mathrm{mag}$ from the PNLF bright cut-off of M105, which is similar to the limiting magnitude of two the ePN.S fields discussed previously.

\subsection{Catalogue matching}

\begin{figure}
  \centering
  \includegraphics[width=8.8cm]{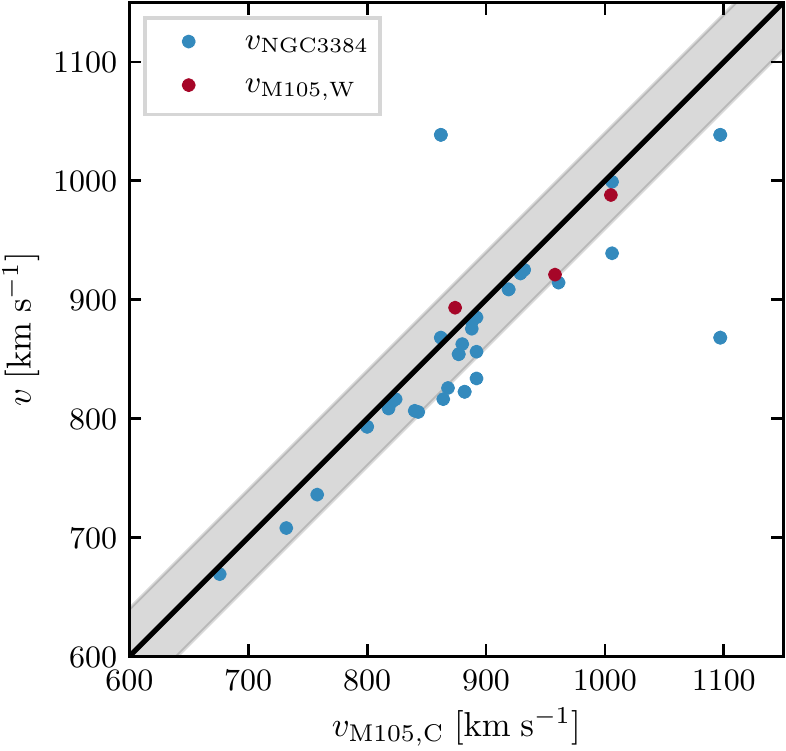}
  \caption[Comparison of PN velocity measurements in different fields of M105.]{Velocity measurements of PNe observed in the central field of M105 compared to those observed in the field centred on NGC~3384 (blue) and the western e$^{2}$PN.S field (red). The grey shaded region indicates the PN.S velocity error of $20\;\mathrm{km}\;\mathrm{s}^{-1}$ .}
  \label{figM105:vel-vel}
\end{figure}

We first identified the PNe that were observed multiple times in the regions of overlap of the PN.S fields (cf. Fig~\ref{figM105:survey}). The largest overlap exists between the fields centred on M105 and NGC~3384, where 30 matching PNe were identified within a matching radius of $5\arcsec$. This seemingly large matching radius was chosen due to the larger positional uncertainties of the PN.S data compared to conventional imaging. We also identified three PNe that were observed both in the field centred on M105 and in the new western e$^{2}$PN.S field. Figure \ref{figM105:vel-vel} shows that the velocity measurements in the different fields agree well. The 11 measurements that lie outside of the grey shaded region that indicates the nominal PN.S velocity error of $20\;\mathrm{km}\;\mathrm{s}^{-1}$ all belong to PNe that were observed close to one of the field edges in either of the two fields. For these PNe, we only considered the measurement taken closer to the respective field centre. For any other PNe with two velocity measurements, we used the mean of these two measurements.
For the remainder of this paper, the name e$^2$PN.S survey encompasses the two fields centred on M105 and NGC~3384 that were originally part of the ePN.S survey, as well as the new fields south and west of M105.

\begin{figure}
    \centering
    \includegraphics[width=8.8cm]{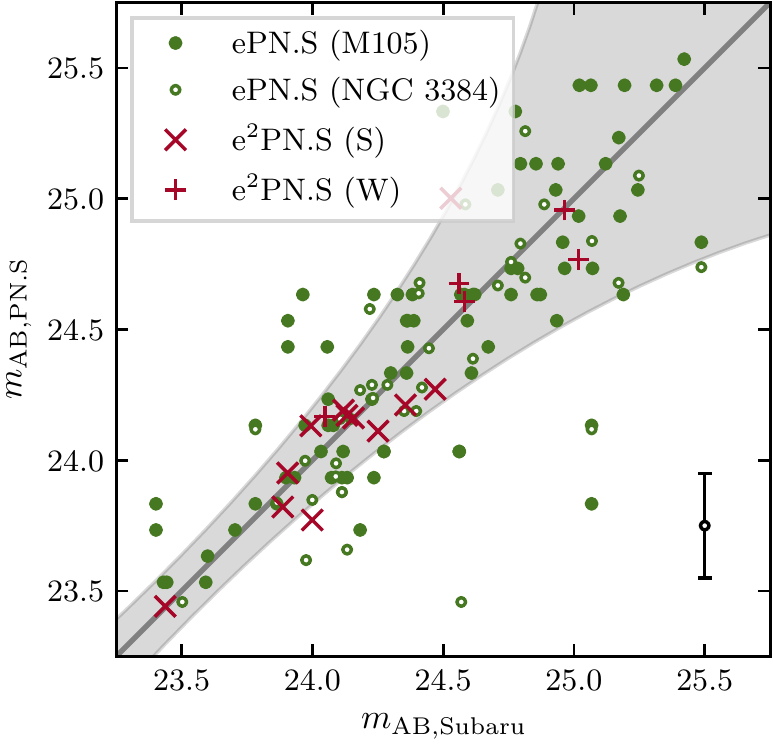}
    \caption[Comparison of PN magnitudes measured with SurpimeCam and the PN.S]{Comparison of magnitudes of PNe measured with the Subaru SuprimeCam and the PN.S. The colour-coding is the same as in Fig.~\ref{figM105:survey}. The grey shaded region indicates where 99\% of Subaru sources would fall if two independent measurements of their magnitudes $m_\mathrm{AB,Subaru}$ were plotted against each other. The error bar in the lower right corner denotes the typical measurement error for magnitudes measured with the PN.S.}
    \label{figM105:mag-mag}
\end{figure}

We then matched the kinematic with the photometric catalogue, again using a matching radius of $5\arcsec$. The unmasked region of overlap between the SuprimeCam and the PN.S fields of view spans $344.2\;\mathrm{arcsec}^2$ on the sky. Figure \ref{figM105:mag-mag} shows the comparison of magnitudes measured from the [\ion{O}{iii}] image obtained with SuprimeCam and from the counter-dispersed images of the PN.S. The grey shaded region indicates where 99\% of Subaru sources would fall if two independent measurements of their magnitudes $m_\mathrm{AB,Subaru}$ were plotted against each other (see Appendix~\ref{app:limmag} for details on the artificial star tests that we carried out). The typical PN.S magnitude errors \citep[$\pm0.2\;\mathrm{mag}$,][]{2007ApJ...664..257D} are indicated by the error bar in the lower right corner of the figure.
In the following section, we use the e$^{2}$PN.S survey as an independent validation of the PN candidates detected with the SuprimeCam photometry.

\subsection{Completeness and sources of contamination in the photometric catalogue}
\label{sec:contamination}

\begin{figure}
    \centering
    \includegraphics{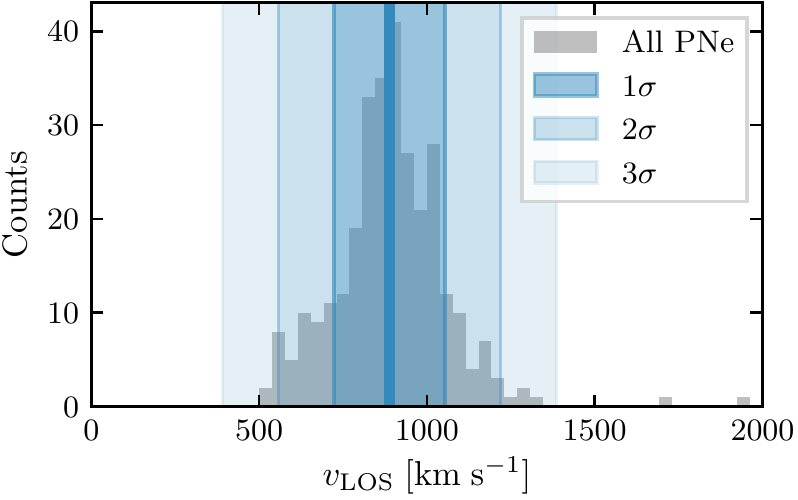}
    \caption{Line-of-sight velocity distribution of PNe from the e$^{2}$PN.S survey. The blue shaded regions denote the $1$, $2$, and $3\sigma$ intervals from the mean velocity (blue vertical line) of the sample. The double-peaked and asymmetric nature of the histogram is due to the fact that the survey covers NGC~3384 as well as M105.}
    \label{fig:pnslosvd}
\end{figure}

\begin{figure}
  \centering
  \includegraphics[width = 8.8cm]{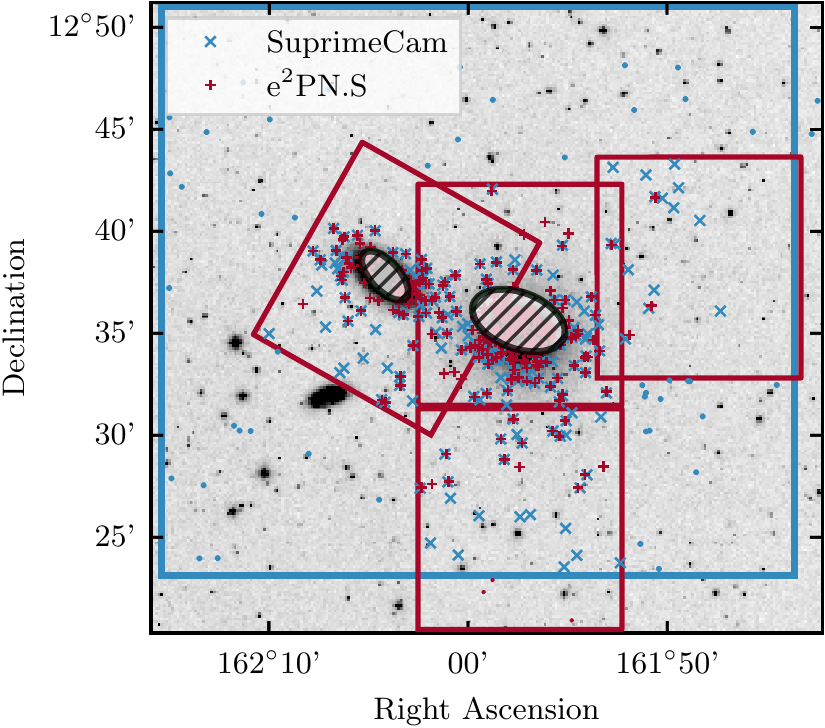}
  \caption{Zoomed-in DSS image centred on M105 with the PNe in the region of overlap between the SuprimeCam (blue rectangle) and e${}^2$ PN.S (red rectangles) surveys marked as blue and red crosses, respectively. The masked-out centres of NGC~3384 and M105 are indicated by hatched ellipses.}
  \label{fig:ly-test-distr}
\end{figure}

\begin{figure}
  \centering
  \includegraphics[width = 8.8cm]{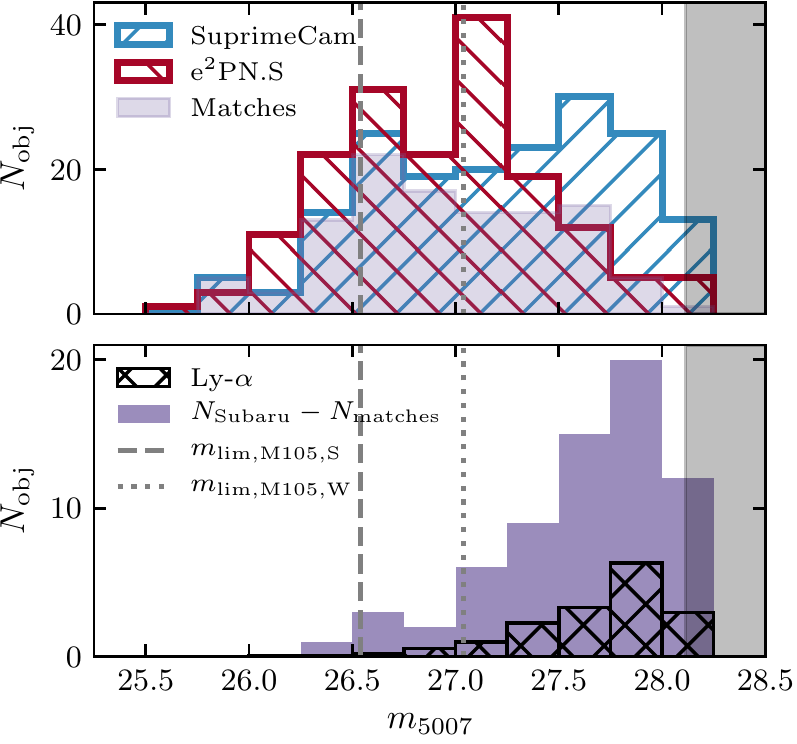}
  \caption{\emph{Top:} Magnitude distribution of PN candidates from the SuprimeCam survey (hatched blue histogram) and PNe from the e${}^2$PN.S survey (hatched red histogram) in the region of overlap highlighted in Fig.~\ref{fig:ly-test-distr}. The magnitude distribution of objects identified in both surveys is shown in purple. The vertical lines denoted the limiting magnitudes of the two shallowest e${}^2$PN.S fields, and the shaded grey region denotes magnitudes fainter than the limiting magnitude of the SuprimeCam survey. \emph{Bottom}: Magnitude distribution of PN candidates from the SuprimeCam survey with no counterpart in the e${}^2$PN.S (purple). The hatched histogram shows the expected distribution of background Ly-$\alpha$ galaxies at redshift $z=3.1$ in the region of overlap scaled by the photometric completeness of the SuprimeCam survey.}
  \label{fig:ly-test}
\end{figure}

The evaluation of the completeness of our catalogue precedes any further quantitative analysis. We determined the \emph{\textup{spatial completeness}} of our SuprimeCam survey by calculating the recovery fraction of the synthetic PN population as a function of radius and the \emph{\textup{photometric completeness}} by calculating the recovery fraction as a function of magnitude (see Appendix~\ref{app:compl} for further details and Tables~\ref{tabM105:spatial} and \ref{tabM105:photometric} for tabulated completeness values).

We furthermore needed to account for the presence of foreground and background contaminants. We estimated that $10\%$ of the PNe candidates might be foreground objects such as faint Milky Way (MW) stars (see Sect.~\ref{app:faint}). This is caused by the so-called spillover effect \citep{2005aj....129.2585a}: due to photometric errors, continuum emitters such as faint foreground stars may end up with measured narrow-band fluxes that are brighter than their intrinsic flux and populate the CMD region from which PN candidates are selected.  Background contaminants might be faint Ly-$\alpha$ galaxies at redshift $z=3.1$ or [\ion{O}{ii}]3727\AA\ emitters at redshift $z = 0.345$, for whose presence we statistically account (see Appendix~\ref{app:lyalpha} for details).  We estimated their fraction to be $24.9\pm5.0\%$ of the completeness-corrected sample.

The PN.S does not cover the second bluer line of the [\ion{O}{iii}] doublet ([\ion{O}{iii}]4959\AA) that is commonly used to distinguish PNe from contaminants in spectroscopic follow-up surveys. It is, however, possible to use the PN.S data to exclude the majority of Ly-$\alpha$ galaxies that are associated with a continuum redder than the [\ion{O}{iii}]5007\AA\ emission line, as the continuum appears as a strike in slitless spectroscopy.

To estimate the contamination from Ly-$\alpha$ galaxies without a measurable continuum, we followed the argument presented by \citet{2018mnras.477.1880s}: the number of background galaxies should be uniformly distributed in a velocity histogram. Figure~\ref{fig:pnslosvd} shows the LOS velocity distribution of PNe from the e$^{2}$PN.S survey, with blue bands denoting the $1$, $2$, and $3\sigma$ intervals from the mean velocity of the sample. The LOS velocity distribution is composed of PNe that are associated with M105 and NGC~3384, which results in a double-peaked asymmetric distribution. From the presence of two PNe in the velocity range $1\,500 - 2\,000\;\mathrm{km}\;\mathrm{s}^{-1}$, beyond $3\sigma$ from the mean velocity, we expect $\text{eight}$ PNe at most in the full velocity range ($0-2\,000\;\mathrm{km}\;\mathrm{s}^{-1}$). Therefore, we estimate the fraction of Ly-$\alpha$ contaminants without a measurable continuum to be 2.6\% at most\  (8 of 307) in the e${}^2$PN.S survey.

We thus used the e${}^2$PN.S survey as an independent validation of the PN candidates detected based on SuprimeCam photometry.
For this, we only considered the region of overlap of the two surveys, as illustrated in Fig.~\ref{fig:ly-test-distr}, and excluded the bright galaxy centres of M105 and NGC~3384 (see Appendix~\ref{app:preprocess} for further details). The top panel of Fig.~\ref{fig:ly-test} shows the magnitude distribution of PN candidates and common objects between the two surveys in this region. The bottom panel of this figure shows the magnitude distribution of PN candidates without a counterpart in the e${}^{2}$ PN.S survey. Within the limiting magnitude of the shallowest e$^{2}$PN.S field ($m^{\star} + 1$), all but one PN candidate have a counterpart in the e$^{2}$PN.S survey. Beyond this limit, the number of PNe candidates without a counterpart increases.
We attribute this to two factors: (1) the larger depth of the SuprimeCam survey and (2) the presence of redshifted Ly-$\alpha$ emitters (black hatched histogram, see Appendix~\ref{app:lyalpha} for details).

\section{Stellar surface photometry of the Leo~I group}
\label{sec:M105SB}

In this section, we provide an overview of the stellar surface photometry of the two brightest galaxies in the field-of-view (FoV) of our PN surveys: M105 and NGC~3384. This allows us to link the properties of the observed PNe to those of the underlying stellar populations in Sect.~\ref{sec:105_alpha}.

\begin{figure*}
  \centering
    \includegraphics[width = 18cm]{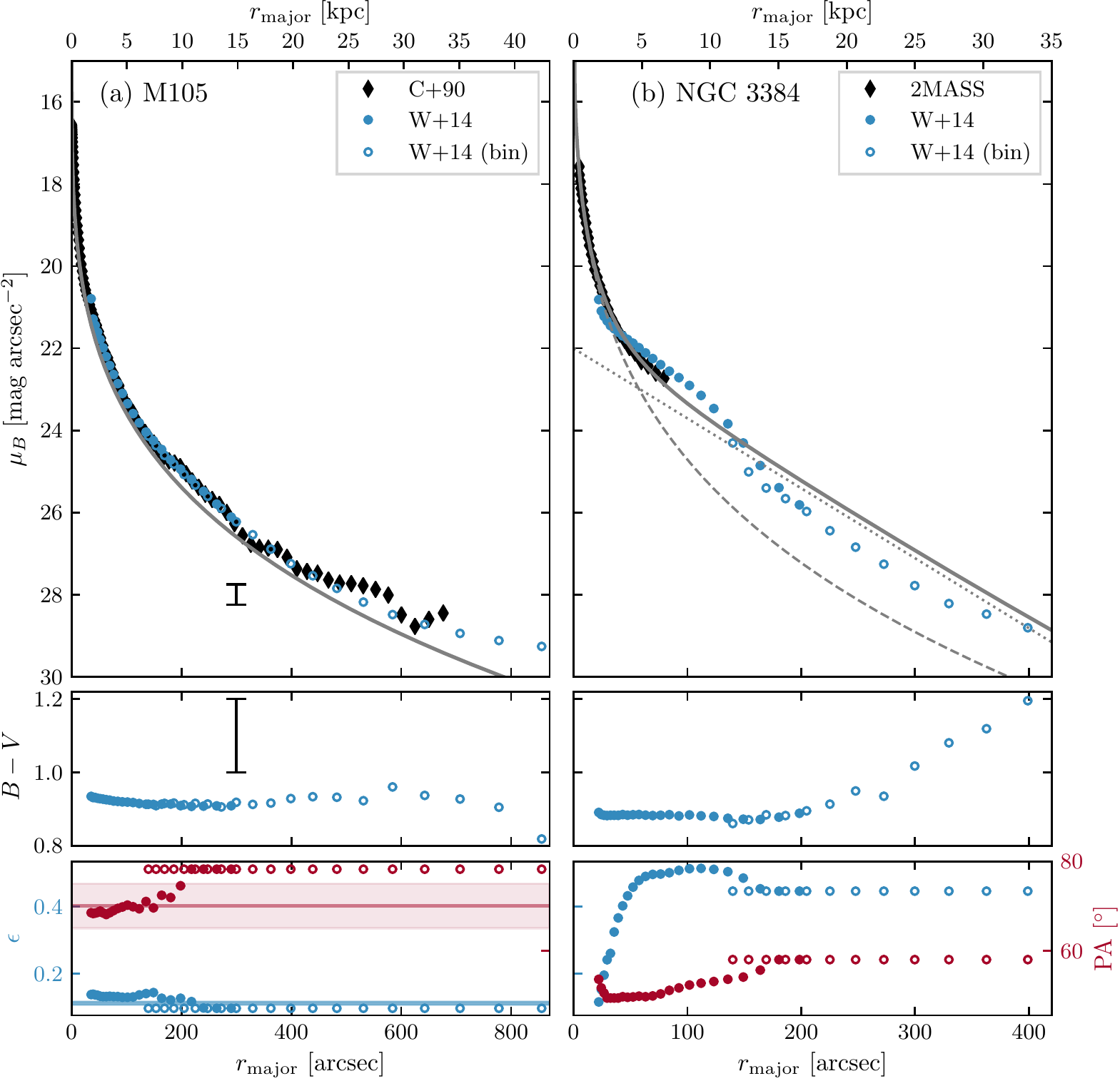}
  \caption{Broad-band photometry of the galaxies M105 (left) and NGC~3384 (right) in the Leo~I group from \citet[][W+14]{2014ApJ...791...38W}. The top panel shows the $B$-band surface brightness profile, the middle panel shows the $B - V$ colour profile, and the bottom panel shows the ellipticity (blue) and position angle (red), all as a
  function of major-axis radius. Open circles denote quantities measured from a binned image (9 pix$\times$ 9 pix). The error bar in the top left panel denotes the typical error on the $B$-band surface brightness, and the error bar the middle panel denotes typical measurement errors on the $B-V$ colour for the binned data. Panel (a) also shows the data from \citet[][black diamonds,]{1990AJ.....99.1813C} and their best-fit de Vaucouleurs profile (grey line) in the top panel. Their mean ellipticity and position angle are indicated by shaded bands in the bottom panel. In panel (b) the
  colour-corrected 2MASS profile \citet[][black diamonds,]{2006AJ....131.1163S}
  and the corresponding disc-spheroid decomposition (dotted, dashed, and solid
  grey lines) by \citet{2013MNRAS.432.1010C} are shown in the top panel.}
  \label{fig:Leophotomery}
\end{figure*}

\subsection{M105}

M105 has been the target of numerous photometric studies: \citet{1948AnAp...11..247D} first analytically described the light distribution of M105, and it was one of the galaxies that were used to define the \citeauthor{1948AnAp...11..247D} law \citep[see also][]{1979ApJS...40..699D}. In the following analyses, we used the relation determined by \citet{1990AJ.....99.1813C} based on $B$-band imaging to describe the surface brightness profile of M105 analytically,
\begin{equation}
  \mu_\mathrm{M105}(r_\mathrm{M105}) = 3.0083 r_\mathrm{M105}^{1/4} + 14.076,
  \label{eqn:m105_SB}
\end{equation}
where $r_\mathrm{M105}$ is the major-axis radius measured from the centre of M105 in units of arcseconds. \citet{1990AJ.....99.1813C} determined the effective radius to be $r_{\mathrm{e,M105}} = 54\farcs8 \pm 3\farcs5$. They furthermore determined the mean ellipticity to be $\epsilon_\mathrm{M105} = 0.111\pm0.005$ and the mean position angle to be $PA = 70\fdg0 \pm 1\fdg0$. These parameters are summarised in Table~\ref{tabM105:galproperties} and the corresponding profiles as a function of radius are shown in Fig.~\ref{fig:Leophotomery}~(a).  \citet{2007MNRAS.379..418C} determined a slightly smaller effective radius $r_{e} = 42\arcsec$ and ellipticity $\epsilon = 0.08$ and a similar photometric position angle ($PA = 67.9\degr$) from \textit{HST}/WFPC2 + MDM $I$-band images. The difference between the effective radii can likely be attributed to the different bands ($I$ band versus $B$ band) in which they were measured and the different limiting surface brightness values of the surveys.

\subsection{NGC 3384}
For completeness, we also briefly discuss the photometric properties of the closest neighbouring galaxy of M105, NGC~3384. Its isophotes overlap those of M105. As NGC~3384 is an S0 galaxy viewed edge-on, its light can be decomposed into a contribution from its bulge (spheroid) and a contribution from its disc. \citet{2013MNRAS.432.1010C} carried out a photometric disc-spheroid decomposition based on 2MASS $K$-band images \citep{2006AJ....131.1163S}. They found that the spheroid light distribution is best described by a S\'{e}rsic profile,
\begin{equation}
    \Sigma_\mathrm{spheroid}(r_\mathrm{3384}) = \Sigma_\mathrm{e}
    \exp\left(-\kappa\left(\left(\frac{r_\mathrm{3384}}{r_\mathrm{e}}\right)^{1/n} - 1\right)\right),
\end{equation}
with $\Sigma_\mathrm{e} = 4.09\times10^{-7}$, $n = 4$, $\kappa = 7.67$, $r_\mathrm{e} = 15\farcs2$, ellipticity $\epsilon_\mathrm{spheroid} = 0.17$, and position angle $PA_\mathrm{spheroid} = 60\fdg51$. The disc light distribution is described by
\begin{equation}
    \Sigma_\mathrm{disc}(r_\mathrm{3384}) = \Sigma_\mathrm{d}\exp\left(-\frac{r_\mathrm{3384}}{r_\mathrm{d}}\right),
\end{equation}
with $\Sigma_\mathrm{d} = 5.01\times10^{-8}$, $r_\mathrm{d} = 63\farcs73$,   $\epsilon_\mathrm{disc} = 0.66$, inclination $i = 70\degr$, and $PA = 52\fdg5$.  These parameters are summarised in Table~\ref{tabM105:galproperties}.

\subsection{Holistic view from a deep, extended multi-band survey}
\label{ssec:multi_band_phot}
As the photometric data described above have been taken at different telescopes and through different filters, their different photometric zero-points had to be taken into account before we were able to use them jointly. We therefore used deep wide-field $B$-band data, encompassing both galaxies, and resulting SB profiles from \citet{2014ApJ...791...38W}. We determined a colour correction of $B-K = -3.75$ for the profiles derived from the 2MASS data by \citet{2013MNRAS.432.1010C}. We furthermore determined a shift of $-0.35\;\mathrm{mag}$ between the $B$-band profiles derived from \citet{2014ApJ...791...38W} and \citet{1990AJ.....99.1813C} and corrected for it accordingly. The colour-corrected $B$-band SB profiles are shown in Figure~\ref{fig:Leophotomery}.

The deeper photometry of \citet{2014ApJ...791...38W} revealed that the surface brightness profile of M105 flattens at large radii with respect to the analytic de Vaucouleurs profile fit by \citet{1990AJ.....99.1813C}. This flattening starts at approximately $750\arcsec$. We discuss the implications of this observation further in Sect.~\ref{sec:M105photmodel}.

\begin{table}
  \centering
  \caption{Properties of M105 and NGC~3384 based on broad-band photometry.}
  \begin{tabular}{lllc}
    \hline
    \hline
    Parameter & Symbol & Value & Ref.\\
    \hline
     & M105 & & \\
    \hline
    SBF distance$^\mathrm{(a)}$  &$D_\mathrm{M105}$ & $10.3$~Mpc & 1 \\
    Effective radius & $r_\mathrm{e,M105}$ & $54\farcs8 \pm 3\farcs5$ & 2 \\
    Mean ellipticity & $\epsilon_\mathrm{M105}$ & $0.111 \pm 0.005$ & 2 \\
    Mean position angle & $PA_\mathrm{M105}$ & $70\degr \pm 1\degr$ & 2 \\
    Mean colour & $(B-V)_\mathrm{M105}$ & $0.92\;\mathrm{mag}$ & 3 \\
    S\'{e}rsic index & $n$ & 4 & 2 \\
    \hline
     & NGC 3384 & & \\
    \hline
    Distance & $D_\mathrm{3384}$ & $11.3$~Mpc & 1 \\
    Mean colour & $(B-V)_\mathrm{3384}$ & $0.88\;\mathrm{mag}$ & 3 \\
    \hline
     &  & Spheroid & \\
    \hline
    S\'{e}rsic index & $n$ & $4$ & 4 \\
    Effective radius & $r_\mathrm{e}$ & $15\farcs2$ & 4 \\
    Mean ellipticity & $\epsilon_\mathrm{spheroid}$ & 0.17 & 4 \\
    Mean position angle & $PA_\mathrm{spheroid}$ & 60\fdg51 & 4 \\
    \hline
     &  & Disc & \\
    \hline
    Disc radius & $r_\mathrm{d}$ & $63\farcs73$ & 4 \\
    Mean ellipticity & $\epsilon_\mathrm{disc}$ & 0.66 & 4\\
    Mean position angle & $PA_\mathrm{disc}$ & $52\fdg5$ & 4\\
    Inclination & $i$ & $70\degr$ & 4 \\
    \hline
  \end{tabular}
  \tablebib{(1)~\citet{2001ApJ...546..681T}; (2)
      \citet{1990AJ.....99.1813C}; (3) \citet{2014ApJ...791...38W}; (4)
      \citet{2013MNRAS.432.1010C}.}
  \tablefoot{
  \tablefoottext{a}{~This SBF distance corresponds to a distance modulus of $\mu_\mathrm{SBF} = 30.049$, which agrees excellently with the distance determined from TRGB magnitudes by
      \citet{2016ApJ...822...70L}, which is $\mu_\mathrm{TRGB} = 30.05 \pm 0.02 \;(\mathrm{random})\; \pm 0.12$ (systematic).}
      }
  \label{tabM105:galproperties}
\end{table}

\section{Radial PN number density profile and the luminosity-specific PN number}
\label{sec:105_alpha}

The luminosity-specific PN number, $\alpha$-parameter for short, relates a PN population with its parent stellar population. The $\alpha$-parameter of a stellar population with $N_\mathrm{PN}$ PNe and total bolometric luminosity $L_\mathrm{bol}$ is
\begin{equation}
  N_\mathrm{PN} = \alpha L_\mathrm{bol}.
\end{equation}
We refer to \citet{2006MNRAS.368..877B} for a theoretical investigation of the relation between the $\alpha$-parameter and the age and metallicity of the underlying stellar population modelled through simple stellar populations (SSPs) and as a function of Hubble type.

\subsection{Radial PN number density profile of M105}
\label{sec:pn_density}
\begin{figure}
  \centering
  \includegraphics[width=8.8cm]{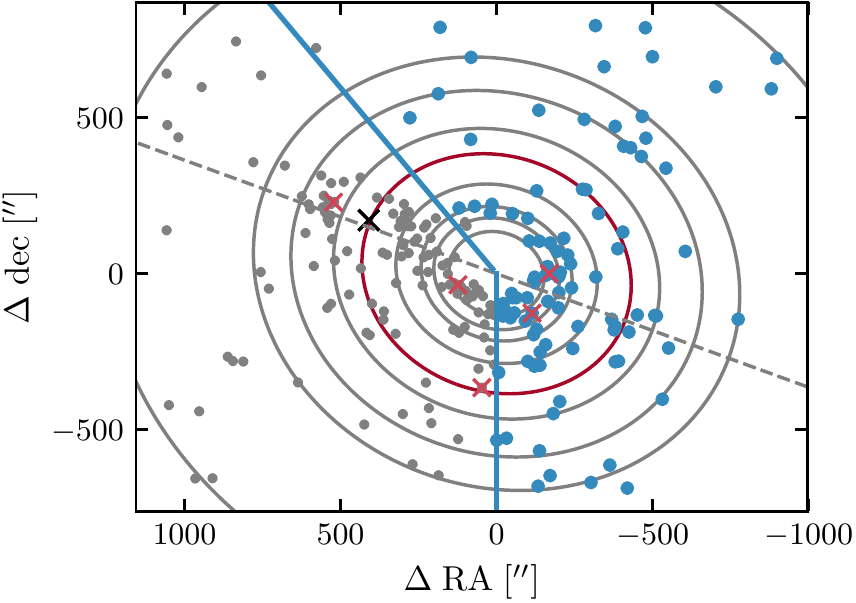}
  \caption{On-sky PN distribution for the PN density and PNLF calculation in M105. Only PNe from the imaging survey and to the right of the blue lines are considered (see text) to limit the contamination from NGC~3384 PNe. The PN density of M105 is calculated in nine elliptical bins denoted in grey. The dashed line indicates the photometric major axis of M105. The red contour corresponds to the vertical dashed line in Fig.~\ref{fig:M105_density}. The black cross denotes the centre of NGC~3384. PNe within 0.5 magnitudes from the PNLF bright cut-off at the distance of M105 are marked with red crosses.
  North is up,  and east is to the left.}
  \label{fig:density_distr}
\end{figure}

In the following analyses, we only considered PNe candidates  from the photometric catalogue from SuprimeCam narrow-band imaging; for this dataset, we could place robust estimates on the spatial and photometric completeness. Furthermore, to limit the contamination from PNe associated with the S0 galaxy NGC~3384, we only considered PNe that are geometrically uniquely associated with M105. As illustrated in Fig.~\ref{fig:density_distr}, we defined two dividing lines (shown in blue), both enclosing $110\degr$ from the negative major axis of M105 and only considered PNe that lie west of these lines (selected PNe are highlighted in blue).

In total, 101 PN were selected for further analysis and were then binned into nine elliptical bins (grey ellipses in Fig.~\ref{fig:density_distr}), whose geometry is defined by the isophotal parameters of M105 as summarised in Table~\ref{tabM105:galproperties}. The bin spacing was chosen such that each bin contained at least 12 PNe. In each bin, we determined the completeness-corrected number of PNe,
\begin{equation}
    N_\mathrm{PN,corr}(r) = \frac{N_\mathrm{PN,obs}(r)\cdot(1 - f_{\mathrm{Ly}\alpha}(r))}{c_\mathrm{s}(r)\cdot c_\mathrm{c}},
\end{equation}
where $c_\mathrm{s}$ and $c_\mathrm{c}$ denote the spatial and colour incompleteness, as detailed in Appendix \ref{app:compl}. The spatial completeness varies with radius and ranges from $77\%$ to $98\%,$ and the average colour completeness of our field is $87\%$.

\begin{figure*}
  \centering
  \includegraphics[width=18cm]{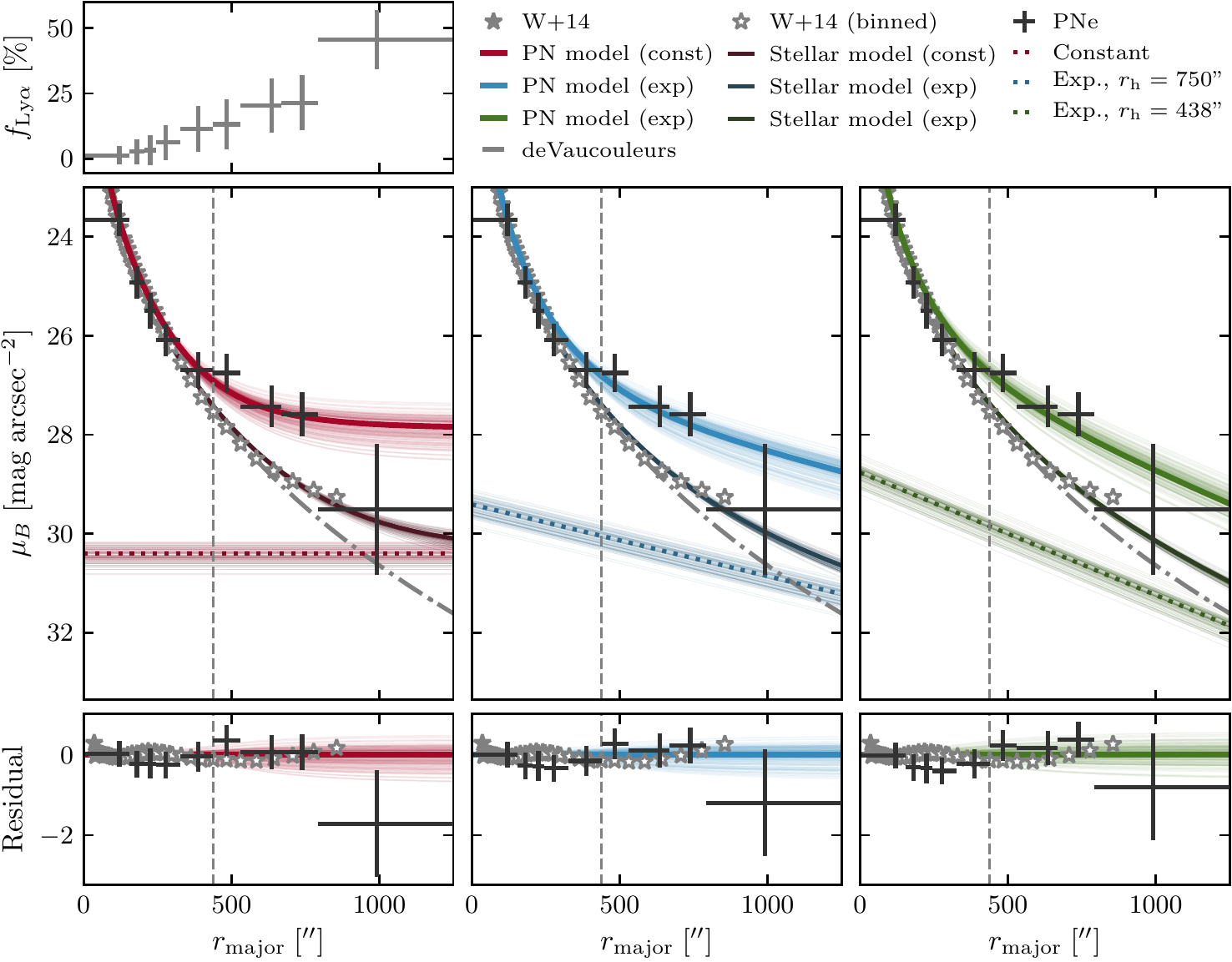}
  \caption{Top panel: Fractional contamination of the PN sample defined in Fig.~\ref{fig:density_distr} by Ly-$\alpha$ emitters in elliptical bins. The three panels of the middle row all show the same data: the grey crosses denote the PN number density calculated from the PN sample defined in Fig.~\ref{fig:density_distr} (101 PNe), which was shifted to the same scale as the $B$-band SB profile \citep[grey stars; open symbols denote binned data,][W+14]{2014ApJ...791...38W} and the bottom panels show the corresponding residuals.  The vertical line denotes the transition from the inner PN-scarce to the outer PN-rich halo based on the change in the PN number density slope.
  In the left column, the data are fit with a two-component SB profile (dark red line) consisting of a de Vaucouleurs-like inner component (dash-dotted grey line) and a constant outer component (dotted red line). The red line denotes the best-fit model to the PN number density.
  In the middle column, the two-component SB profile (dark blue line) also has a de Vaucouleurs-like inner component (dash-dotted grey line), but the outer component is an exponential profile with scale height $r_h = 750\arcsec$ (dotted blue line). The blue line denotes the best-fit model to the PN number density.
  In the right column, the two-component SB profile (dark green line) has a de Vaucouleurs-like inner component (dash-dotted grey line) and an exponential component with scale height $r_h = 438\arcsec$ (dotted green line). The green line denotes the best-fit model to the PN number density.
  The many thin lines are indicative of error ranges for the modelled profiles and were determined with Monte Carlo techniques (cf. Appendix~\ref{app:posterior}).}
  \label{fig:M105_density}
\end{figure*}

The expected luminosity-averaged fraction of Ly-$\alpha$ emitters at redshift $z=3.1$ is $f_{\mathrm{Ly}\alpha} = 24.9\pm5 \%$, averaged over the SuprimeCam FoV (see Appendix~\ref{app:lyalpha}). The distribution of Ly-$\alpha$ contaminants can be approximated to be homogeneous in the SuprimeCam FoV. This implies that their contribution becomes stronger in the outer regime of low SB and low-PN number density. We therefore estimated the relative contribution of Ly-$\alpha$ emitters in each of the elliptical bins illustrated in Fig.~\ref{fig:density_distr}: we randomly distributed the expected number of Ly-$\alpha$ emitters in the unmasked survey area and determined how many fell in each bin. We repeated this process $10\,000$ times to estimate error bars. The result is shown in the top panel of Fig.~\ref{fig:M105_density}.

The PN logarithmic number density profile is then
\begin{equation}
    \mu_\mathrm{PN}(r) =
-2.5\log_{10}\left(\frac{N_\mathrm{PN,corr}(r)}{A(r)}\right) + \mu_\mathrm{off},
    \label{eqn:density}
\end{equation}
where $A(r)$ is the area of intersection of the elliptical annulus with the FOV of the observation, as illustrated in Fig.~\ref{fig:density_distr}. From the offset $\mu_\mathrm{off}$ to the observed stellar SB profile, the $\alpha$-parameter can be determined (see Sect.~\ref{sec:M105photmodel}).

The PN number density profile shown in Fig.~\ref{fig:M105_density} was shifted to be on the same scale as the $B$-band surface brightness profile from \citet{2014ApJ...791...38W}. Within $\sim 438\arcsec (\approx 9 r_\mathrm{e})$ from the centre of M105, PN number density and stellar SB agree very well with each other, but beyond this transition radius, the PN number density profile significantly flattens with respect to the stellar SB profile. We note that the flattening occurs at $250\arcsec$ shorter radial distance compared to the flattening of the SB profile that is observed beyond $\sim 750\arcsec$. As already discussed in \citet{2013A&A...558A..42L, 2015A&A...579A.135L}, we do not expect the flattening to be due to contaminants such as Ly-$\alpha$-emitting galaxies because their contribution is statistically subtracted. The origin of the flattening of the PN number density is investigated in the following section.

\subsection{Two-component photometric models for the extended halo of M105}
\label{sec:M105photmodel}

The flattening of the PN number density profile at large radii may be caused by a diffuse outer component. The second component,  whether an additional component of the extended halo or in the form of diffuse  IGL, is corroborated by the flattening of the stellar SB profile observed by \citet{2014ApJ...791...38W}. Such a flattening of the PN number density profile was also observed in the outer regions of other ETGs at the centres of groups and clusters of galaxies, such as M87 in the Virgo cluster  \citep{2013A&A...558A..42L, 2015A&A...579A.135L, 2018A&A...620A.111L} and M49  in the Virgo subcluster B \citep{2017A&A...603A.104H, 2018A&A...616A.123H}. It was attributed to intra-cluster light (ICL) or IGL with a  distinct $\alpha$-parameter value and spatial density distribution from the main halos of the galaxies.

To reproduce the flattening of the stellar SB profile at large radii, we constructed two sets of two-component photometric models, where the two components are denoted \emph{\textup{"inner"}} and \emph{\textup{"outer"}} according to the respective radial ranges in which they dominate the observed PN number density and SB profiles.
In both sets of models, the stellar light distribution of the inner halo was modelled with a de Vaucouleurs profile as fitted by \citet{1990AJ.....99.1813C}. We expressed it in terms of a S\'{e}rsic profile \citep{1963baaa....6...41s},
\begin{equation}
    \mu_\mathrm{inner}(r) = \mu_\mathrm{e} + k(n)\left(\left(\frac{r}{r_\mathrm{e}}\right)^{1/n} - 1 \right),
    \label{eqn:sersic}
\end{equation}
with
\begin{equation}
    k(n) = 2.17n - 0.355.
\end{equation}
Equation~\eqref{eqn:sersic} is equivalent to Eq.~\eqref{eqn:m105_SB} with a S\'{e}rsic index $n = 4$, $r_\mathrm{e} = 58\farcs6$, and $\mu_\mathrm{e} = 22.05\;\mathrm{mag}\,\mathrm{arcsec}^{-2}$, after the colour correction determined in Sect.~\ref{ssec:multi_band_phot} is applied.

For the outer component, we considered two models:
\begin{enumerate}
  \item A constant flat SB profile $\mu_\mathrm{outer}(r) = \mu_\mathrm{const}$, which was also used by \citet{2013A&A...558A..42L} and \citet{2017A&A...603A.104H} as part of their two-component models of the SB and PN number density profiles of galaxies in group and cluster environments.
  \item An exponential profile defined as
  \begin{equation}
      \mu_\mathrm{outer}(r) = \mu_0 + 1.086 \left(\frac{r}{r_\mathrm{h}}\right),
  \end{equation}
  where $\mu_0$ is its central surface brightness and $r_\mathrm{h}$ its scale height, as used by \citet{2016ApJ...820...42I}, for example, to describe the extended halo of NGC~1399 at the core of the Fornax cluster.
\end{enumerate}
The total SB of the two-component models is then given by
\begin{equation}
    \mu_\mathrm{total}(r) = -2.5 \log_{10}\left(10^{-0.4\mu_\mathrm{inner}(r)} + 10^{-0.4\mu_\mathrm{outer}(r)}\right).
    \label{eq:M105stellarmodel}
\end{equation}
In order to account for the discrepancy between the observed PN number density and stellar SB profiles at large radii (beyond $438\arcsec$), we assumed that the two components have different $\alpha$-parameters \citep{2013A&A...558A..42L,2017A&A...603A.104H}:
\begin{align}
        \tilde{\Sigma}(r) &= (\alpha_{\Delta m, \mathrm{inner}}
        I_{\mathrm{inner,bol}}(r) + \alpha_{\Delta m, \mathrm{outer}}
        I_{\mathrm{outer,bol}}(r))s^2 \nonumber \\
        &= \alpha_{\Delta m,
        \mathrm{inner}} \left(I_{\mathrm{inner,bol}}(r) +
        \left(\frac{\alpha_{\Delta m, \mathrm{outer}}}{\alpha_{\Delta m,
        \mathrm{inner}}} - 1 \right) I_{\Delta m,
        \mathrm{outer,bol}}(r)\right)s^2 . \label{eq:M105photmodel}
\end{align}
The two components $I_{\mathrm{inner,bol}}(r)$ and $I_{\mathrm{outer,bol}}(r)$ are the bolometric SB profiles weighted by the $\alpha$-parameters of the respective populations, and $s = 49.6\;\mathrm{pc}/\arcsec$ is the conversion factor from angular to physical units for a distance of $D = 10.3\;\mathrm{Mpc}$.

The bolometric SB profiles $I_\mathrm{bol}(r)$ were obtained from the observed SB profiles $\mu(r)$:
\begin{equation}
        I(r) = 10^{-0.4(\mathrm{BC}_B - \mathrm{BC}_{\odot})}10^{-0.4(\mu(r) - K)},
          \label{eq:bolo}
\end{equation}
with the solar bolometric correction $\mathrm{BC}_{\odot} = -0.69$, and $K = 26.98\,\mathrm{mag}\,\mathrm{arcsec}^{-2}$ is the $B$-band conversion factor to physical units $L_{\odot}\,\mathrm{pc}^{-2}$ \citep{2006MNRAS.368..877B}.
While a fixed value of $\mathrm{BC}_V = -0.85$ with 10\% accuracy can be assumed based on the study of stellar population models for different galaxy types, the bolometric correction in the $B$-band depends on galaxy colour \citep{2006MNRAS.368..877B},
\begin{equation}
    \mathrm{BC}_B = \mathrm{BC}_V - (B - V)_\mathrm{gal}.
\end{equation}
Assuming a colour of $B - V = 0.92$ (cf. middle panel of
Fig.~\ref{fig:Leophotomery}a), we used a value of $\mathrm{BC}_B = -1.76$.

Instead of fitting $\alpha_\mathrm{inner}$ and $\alpha_\mathrm{outer}$ simultaneously, we first determined the well-constrained $\alpha$-parameter of the inner component from the offset $\mu_\mathrm{off}$ between the PN number density and the inner stellar SB profile. Due to the large depth of our survey, we can directly determine the commonly used $\alpha_{2.5}$ parameter, which is calculated in a magnitude range from $m^{\star}$ to $m^{\star} + 2.5$. In this magnitude range, we found the offset to the stellar SB profile to be $\mu_\mathrm{off} = 16.7 \pm 0.1 \;\mathrm{mag}\;\mathrm{arcsec}^{-2}$.
When the bolometric correction is applied as defined in Eq.~\eqref{eq:bolo}, this offset translates into an $\alpha$-parameter value of $\alpha_\mathrm{2.5,inner} = (1.17 \pm 0.10)\times 10^{-8}\;\mathrm{PN}\;L^{-1}_\mathrm{bol}$.

\citet{2006MNRAS.368..877B} determined $\alpha_\mathrm{8,M105} = (1.69\pm0.12)\times10^{-7}\;\mathrm{PN}\;L^{-1}_\mathrm{bol}$ based on the PN.S pointing centred on M105. We scaled this value from $\alpha_8$ to $\alpha_{2.5}$ using the standard PNLF from \citet{1989ApJ...339...53C}, resulting in $\alpha_\mathrm{2.5, M105} = (1.66\pm0.12)\times10^{-8}\;\mathrm{PN}\;L^{-1}_\mathrm{bol}$. This value is similar to the value determined in this work, but slightly higher. The difference may stem from a different treatment of the completeness corrections for the PN.S and SuprimeCam data.
\begin{table*}[]
\centering
\caption{Best-fit parameters for analytic models fit to the SB and PN number density.}
\begin{tabular}{lllllll}
\hline
\hline
Model & Fit parameters & & & & BIC & $\chi^2_\mathrm{red}$ \T \\
S\'{e}rsic +  & $\mu_\mathrm{off}\;[\mathrm{mag}\,\mathrm{arcsec}^{-2}]$ & $\mu_0\;[\mathrm{mag}\,\mathrm{arcsec}^{-2}]$ & $r_\mathrm{h}\;[\arcsec]$ & $\alpha_{2.5, \mathrm{outer}}/\alpha_{2.5,\mathrm{inner}}$ & & \B \\ \hline
Constant & $30.4 \pm 0.1$ & -- & -- & $11.3 \pm 4.8$ & $48$  & $2.28$ \T \\
Exponential 1 & -- & $29.4 \pm 0.1$ & $750$ (fixed) & $10.0 \pm 4.8$ & $63$  & $2.99$            \\
Exponential 2 & -- & $28.8 \pm 0.2$ & $438$ (fixed) & $9.4 \pm 4.9$  & $73$  & $3.69$ \\
\hline
\hline
Model & Fit parameters & & & & BIC & $\chi^2_\mathrm{red}$ \T \\
 & $\mu_\mathrm{eff}\;[\mathrm{mag}\,\mathrm{arcsec}^{-2}]$ & $n$ & $r_\mathrm{eff}\;[\arcsec]$ & $\alpha_\mathrm{2.5,inner}\;[\times 10^{-8}\;\mathrm{PN}\;L^{-1}_\mathrm{bol}]$ & & \B \\
\hline
Single S\'{e}rsic & $22.3\pm0.1$ & $4$ (fixed) & $66.2\pm2.2$ & $1.57\pm0.49$ & $107$ & $6.62$ \T \\
\hline
\hline
\end{tabular}
\label{tab:fit_res}
\end{table*}

We then proceeded to fit the outer SB profiles and corresponding $\alpha$-parameters for the exponential and constant models using the non-linear least-squares minimisation routine \textsc{lmfit} \citep{Newville_2014_11813}. We jointly minimised the residuals of the total stellar SB and PN number density profiles. In order to compare the models, we determined the Bayesian information criterion (BIC) as well as the reduced $\chi^2_\mathrm{red}$ for each of the best-fit models. Table~\ref{tab:fit_res} summarises the best-fit parameters, their error margins, and the goodness-of-fit parameters. For completeness, we also show the results of the fit of a single S\'{e}rsic profile to the data. We fixed the S\'{e}rsic index to $n=4$ because otherwise, it was not possible to constrain the effective radius $r_\mathrm{e}$ and central surface brightness $\mu_\mathrm{eff}$.
This model is ruled out due to its high BIC and reduced $\chi^2_\mathrm{red}$ values compared to the two-component models discussed in the previous section.

For the first model, which has the constant outer SB profile, the best-fit parameters were $\mu_\mathrm{const} = 30.4 \pm 0.1$ and the ratio of the $\alpha$-parameters $\alpha_{2.5, \mathrm{outer}}/\alpha_{2.5,\mathrm{inner}} = 11.3 \pm 4.8$. The central and bottom left panels of Fig.~\ref{fig:M105_density} show the best-fit stellar SB (dark red) and PN number density profiles (red) in comparison to the data and their residuals. We drew 100 random samples for each free parameter from their posterior probability distribution to illustrate the error margins of the total model and its components (see Appendix~\ref{app:posterior} for a detailed description).

For the second model, which has an exponential outer SB profile, we fit the free parameters were the central SB $\mu_0$, the ratio of the $\alpha$-parameters $\alpha_{2.5, \mathrm{outer}}/\alpha_{2.5,\mathrm{inner}}$, and the scale height of the exponential profile $r_\mathrm{h}$. However, the latter proved to be very hard to constrain with the current data. We therefore decided to explore two models: one (Exponential 1) with a fixed $r_h = 750\arcsec$, which is the major-axis transition radius at which the stellar SB profile deviates from a single de Vaucouleurs profile, and a second model (Exponential 2) with a fixed $r_h = 438\arcsec$, which is the major-axis radius at which the PN number density profile deviates from the de Vaucouleurs profile. We thus only fit $\mu_0$ and $\alpha_{2.5, \mathrm{outer}}/\alpha_{2.5,\mathrm{inner}}$, whose best-fit values are summarised in Table~\ref{tab:fit_res}. The resulting best-fit SB and PN number density profiles are shown in Fig.~\ref{fig:M105_density}.

Based on the goodness of fit parameters BIC and $\chi^2_\mathrm{red}$, the best-fit model has a constant outer SB profile because it has the lowest values. However, an outer component with constant SB and stars on radial orbits would not be compliant with the Jeans equations and a positive distribution function. With the current PN data, we cannot constrain the functional form of the component contributing to the stellar SB and PN number density profiles at large radii. As a next step, we used the constraints from the observed distribution of resolved MP and MR stellar populations in M105, which we describe and address in the next section.

\section{Constraints from resolved stellar populations}
\label{sect:RGB}
\begin{table*}[]
\centering
\caption{Best-fit parameters for analytic models fit to the SB and PN number density, with additional constraints from the RGB profiles from \citet{2016ApJ...822...70L}. The overall best-fit profile is emphasised in \textbf{bold}.}
\begin{tabular}{lllllll}
\hline
\hline
Model & Fit parameters & & & & BIC & $\chi^2_\mathrm{red}$ \T \\
\citet{2016ApJ...822...70L}& $\alpha_\mathrm{2.5,MR}\;[\times 10^{-8}\;\mathrm{PN}\;L^{-1}_\mathrm{bol}]$ & $\alpha_{2.5, \mathrm{MP}}/\alpha_{2.5,\mathrm{MR}}$ & & &  \B \\
\hline
Double S\'{e}rsic & $1.25\pm0.08$ & $2.79\pm0.91$ & & & $119$ & $7.81$ \T \\
\hline
\hline
\textbf{Model} & Fit parameters & & & & BIC & $\chi^2_\mathrm{red}$ \T \\
Double S\'{e}rsic + & $\mu_0\;[\mathrm{mag} \mathrm{arcsec}^{-2}]$& $r_\mathrm{h}\;[\arcsec]$ & $\alpha_{2.5,\mathrm{MR+IM}}\;[\times 10^{-8}\;\mathrm{PN}\; L^{-1}_\mathrm{bol}]$ & $\alpha_{2.5, \mathrm{MP}}/\alpha_{2.5,\mathrm{MR+IM}}$ & & \B \\
\hline
Exponential & $27.7\pm0.1$ & $358\pm2$ & $1.00\pm0.11$ & $7.10 \pm 1.87$ & $2$ & $0.92$ \\
\hline
\hline
\end{tabular}
\label{tab:fit_rgb}
\end{table*}

\begin{figure*}
  \includegraphics[width=8.8cm]{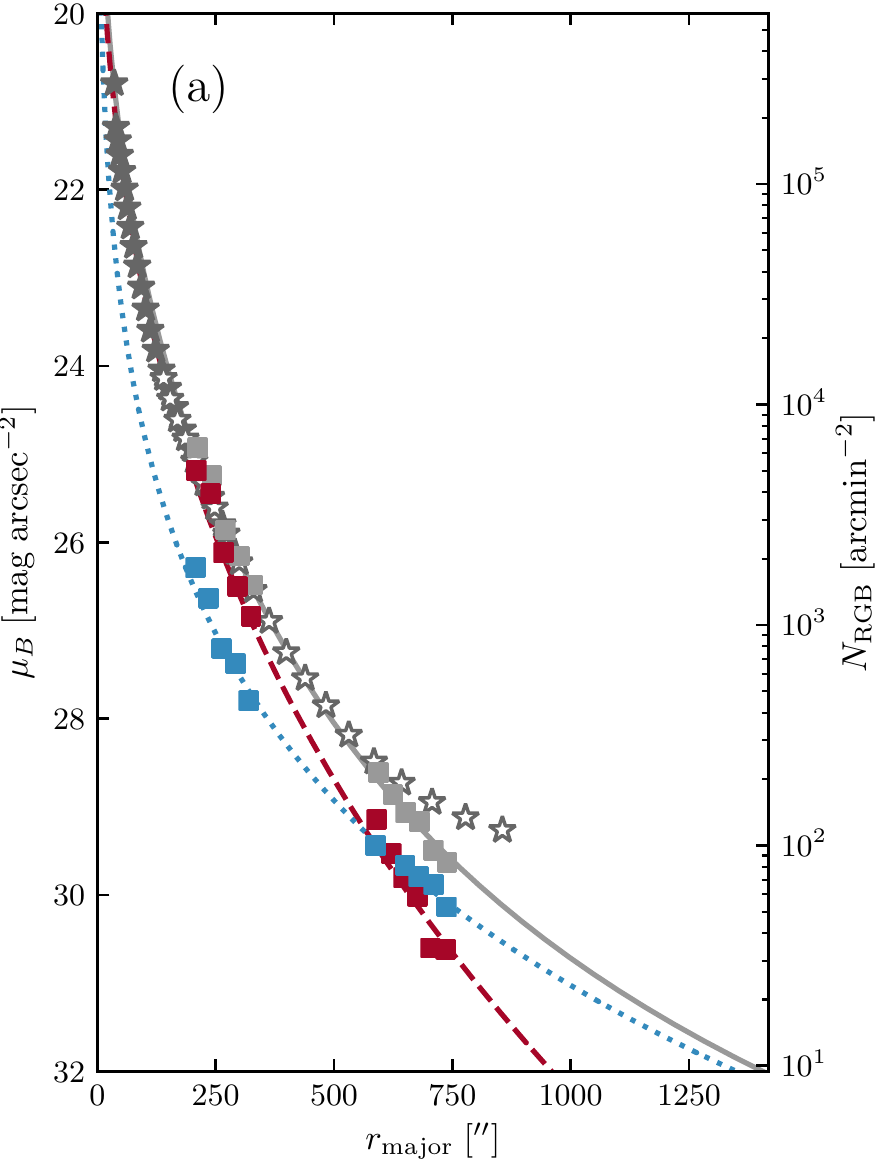}
  \includegraphics[width=8.8cm]{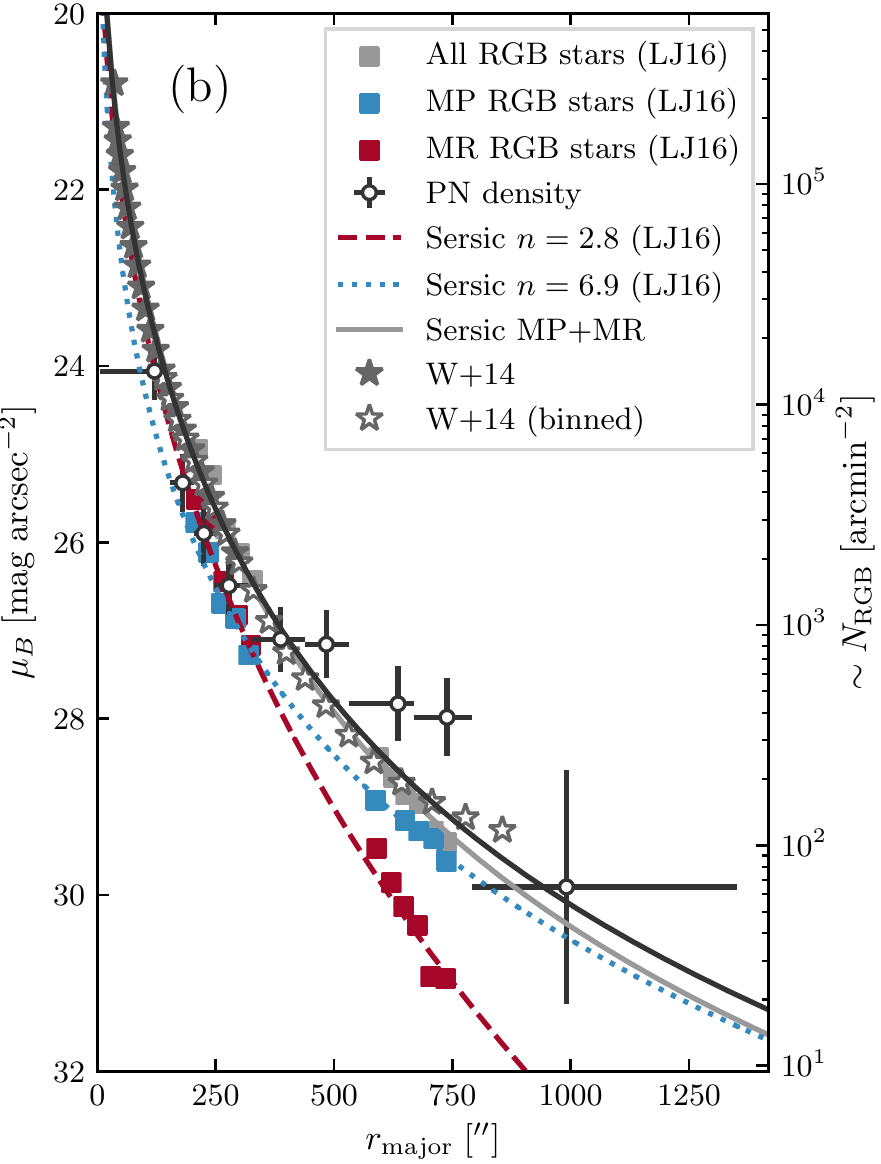}
  \caption{(a) Comparison of the integrated $B$-band SB profiles \citep[grey stars; open symbols denote binned data][]{2014ApJ...791...38W} with resolved stellar population studies of RGB stars \citep{2016ApJ...822...70L}: grey squares indicate the number densities of all RGB stars, blue squares MP ([M/H] $< 0.7$) stars and red squares MR ([M/H] $\geq 0.7$) stars. The coloured lines denote the best-fit S\'{e}rsic profiles to the respective stellar population determined by \citet{2016ApJ...822...70L} with the best-fit indices stated in the legend of panel (b). The RGB number densities were converted into SB using a constant conversion factor. (b) Same as panel (a), but with a metallicity-weighted conversion factor from RGB number densities to SB (see Sect.~\ref{ssec:rgb2sb}) The right $y$-axis is only an indication of the RGB number density, as the conversion from $\mu_B$ is metallicity dependent. In addition, we also show the PN number density (black crosses). The solid black line represents the modelled PN number density profile, where the model consists of the MP and MR S\'{e}rsic profiles weighted by the respective $\alpha$-parameter values.}
  \label{fig:RGB_comp}
\end{figure*}

In order to explain the excess of PNe in the outer halos of early-type galaxies such as M49 and M87 in the Virgo cluster, \citet{2017A&A...603A.104H,2018A&A...616A.123H} and \citet{    2015A&A...579A.135L, 2018A&A...620A.111L} invoked the presence of an accreted MP population at large radii with a higher $\alpha$-parameter value than that of the stellar population at the inner regions of these galaxies.
In addition to the higher $\alpha$-parameter value, further evidence for the presence of these MP stellar populations in the outer regions of M49 and M87 came from a change in PNLF slope at large radii, the gradients towards bluer colours, and from the kinematics of the PNe.

In the case of M105, the direct evidence for an old halo population of MP stars comes from HST photometry \citep{2007ApJ...666..903H,    2016ApJ...822...70L}. The grey hashed squares in Fig.~\ref{figM105:survey} indicate the locations of the two HST pointings in the halo of M105. In Fig.~\ref{fig:RGB_comp}~(a) and (b), we show the integrated light profile \citep[grey stars,][]{2014ApJ...791...38W}, the PN number density (black crosses), and the total number density of RGB stars (grey squares), as well as the number density profiles of the MP (blue squares) and MR RGB stars (red squares) and the corresponding S\'{e}rsic profile fits \citep{2016ApJ...822...70L}. As was emphasised by \citet{2007ApJ...666..903H}, we note that the number density of MP stars exceeds that of the MR stars at approximately $\sim 12\;R_\mathrm{eff}$, at a similar radius where the photometric SB profile deviates from a single S\'{e}rsic profile.

\subsection{Comparing RGB number density and photometric SB profiles}
\label{ssec:rgb2sb}
In order to compare the number density profiles from RGB stars \citep{2016ApJ...822...70L} with the photometric SB profile \citep{2014ApJ...791...38W}, \citet{2016ApJ...822...70L} determined a constant conversion factor from the star counts in the inner HST field ($200\arcsec - 400\arcsec$). However, using this constant conversion factor leads to a discrepancy between the modelled $B$-band SB at large radii and the SB measured by \citet{2014ApJ...791...38W}. Fig.~\ref{fig:RGB_comp}(a) shows that the measured photometric SB profile is brighter than the SB inferred from the RGB number counts. This discrepancy cannot be attributed to the measurement errors on the $B$-band SB, which are about $\pm0.25~\mathrm{mag}$ for the binned data from \citet{2014ApJ...791...38W}. The origin must reside somewhere else: we hypothesise is that it is linked to the different contribution to the SB profile from the MR and MP stars as a function of radius.

A MP population of RGB stars is expected to yield a higher $B$-band luminosity than a MR population. We thus used the stellar population models from the MILES stellar library \citep{2010MNRAS.404.1639V} to estimate the luminosities emitted by the stellar populations with ages and metallicities that were determined for the MP and MR RGB stars by \citet{2016ApJ...822...70L} in the different regions in M105. We used BaSTi isochrones \citep{2013A&A...558A..46P} and a \citet{2003PASP..115..763C} initial mass function. The ages of both populations were set to 12~Gyr. The peak of the metallicity distribution function (MDF) of the MR RGB stars is [M/H]$\approx0.0,$ and the peak metallicity of the MDF of the MP RGB stars is [M/H]$\approx-1.1$ \citep[][Fig.~7]{2016ApJ...822...70L}. The resulting scaling factors for the S\'{e}rsic profiles fit by \citet{2016ApJ...822...70L} are then 1.42 for the MP profile and 0.81 for the MR profile. The corrected data points and S\'{e}rsic profiles are shown in Fig.~\ref{fig:RGB_comp}(b), which now agree better with the SB profile from \citet{2014ApJ...791...38W}.

\subsection{Multi-component photometric models}
\label{ssec:final_model}

\begin{figure}
  \includegraphics[width=8.8cm]{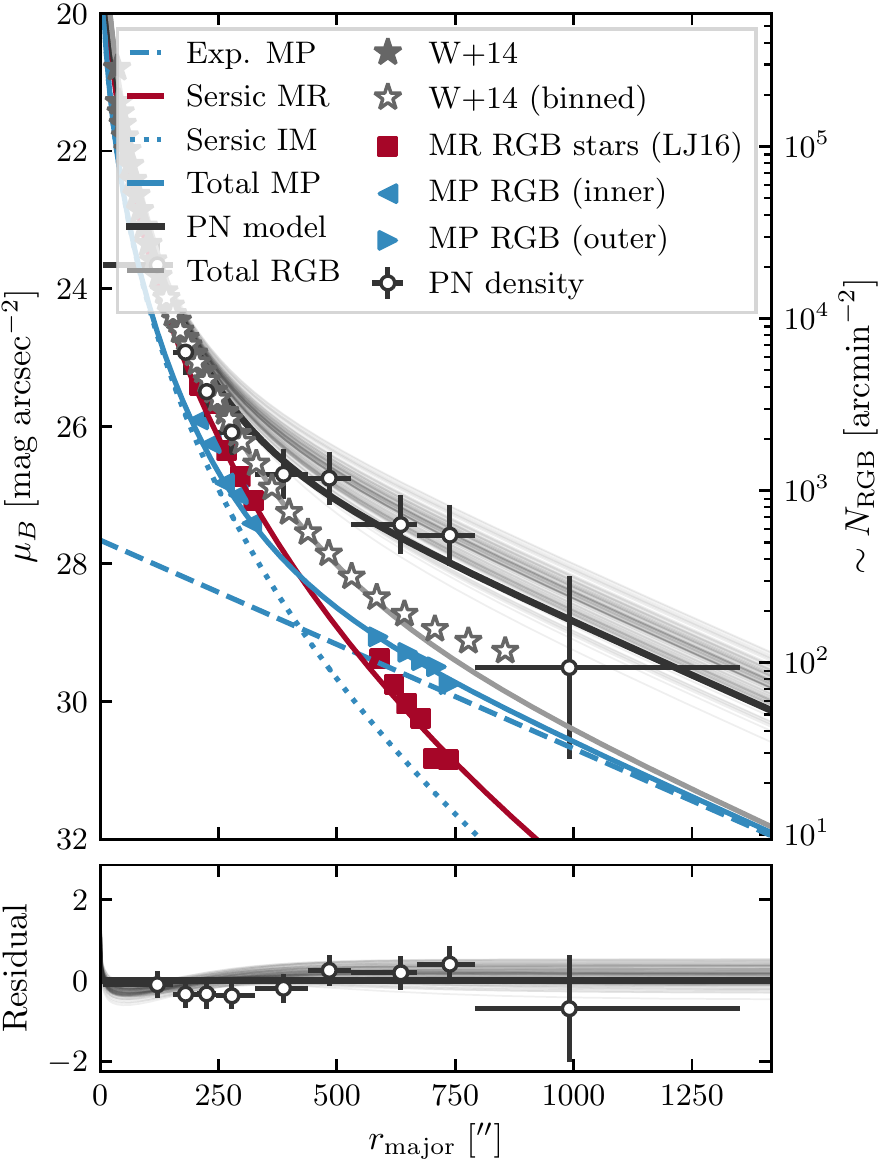}
  \caption{Final stellar population decomposition of the integrated SB and PN number density profiles. The top panel shows the final best-fit model (black line) to the PN number density (black crosses). Comparison of the integrated $B$-band SB profiles \citep[grey stars; open symbols denote binned data][]{2014ApJ...791...38W} with resolved stellar population studies of RGB stars \citep{2016ApJ...822...70L}: blue triangles indicate MP stars, and red squares show MR stars. We used a metallicity-weighted conversion factor from RGB number densities to SB (see Sect.~\ref{ssec:rgb2sb}). The right $y$ axis is thus only an indication of the RGB number density. The RGB number density profile is decomposed into an exponential MP profile (dashed blue line), an intermediate-metallicity (IM) S\'{e}rsic profile (dotted blue line), and the MR S\'{e}rsic (solid red line). The many thin grey lines are indicative of error ranges and were determined using Monte Carlo techniques (cf. Appendix \ref{app:posterior}). The lower panel shows the residuals of the best fit.}
  \label{fig:density_final}
\end{figure}

Following the method presented in Sect.~\ref{sec:M105photmodel}, we then explored whether the PN number density profile may be reproduced as the sum of the two S\'{e}rsic profiles previously fitted to the MP and MR RGB stars distribution by \citet{2016ApJ...822...70L}, allowing for a different $\alpha$-parameter value for each population. The metallicity bins defined by \citet{2016ApJ...822...70L} are [M/H] $< -0.7$ for the MP and [M/H] $\geq 0.7$ for the MR RGB stars. We first calculated the $\alpha$-parameter value for the inner halo, where the MR RGB stars dominate, and kept all the parameters of the S\'{e}rsic profile determined by \citet{2016ApJ...822...70L} fixed.
This resulted in $\alpha_\mathrm{MR} = (1.25\pm0.08)\times 10^{-8} \;\mathrm{PN}\;L^{-1}_\mathrm{bol}$, which agrees well with the $\alpha$-parameter value determined for the inner halo of M105 in Sect.~\ref{sec:M105photmodel}.
We then determined the $\alpha$-parameter value of the MP population by fitting the ratio $\alpha_\mathrm{MR}/\alpha_\mathrm{MP}$ as shown in Fig.~\ref{fig:RGB_comp}~(b). The fit to the PN number density results in $\alpha_\mathrm{MR}/\alpha_\mathrm{MP} = 2.79 \pm 0.91$. However, the corresponding PN number density model is not a good representation of the measured PN number density profiles. It slightly overestimates the number of PNe in the inner halo and significantly underestimates the number of PNe at large radii. This is also reflected in the goodness-of-fit metrics reported in Table~\ref{tab:fit_rgb}.

We then examined the MDFs in the two HST fields, and observed that while the RGB population with metallicity $-1.0 \le \mathrm{[M/H]} < -0.5$ follows the S\'{e}rsic profile of the MR RGB population $(\mathrm{[M/H]} \ge -0.5),$ the very MP RGB population with [M/H] $< -1.0$ fractionally increases the most in the outer halo.
We therefore adopted the following alternative assumptions that
\begin{enumerate}
    \item the intermediate metallicity ($-1.0 \le \mathrm{[M/H]} < -0.5$) and MR RGB stars in the inner halo of M105, that is, those in the south-eastern HST pointing, can be fitted by a single number density profile with  slope $n = 2.8$ (as fitted fit to the MR stars in the inner and outer halo by \citet[][right panel of Fig.~8 therein]{2016ApJ...822...70L}), denoted by the blue dotted line in Fig.~\ref{fig:density_final},
    \item there is a distinct additional component of MP RGB stars ([M/H] $< -1.0$ ), which accounts for the higher fraction of MP stars at large radii, that is, in the western HST field.
\end{enumerate}

The MDFs fitted by \citet[][Table 4 therein,]{2016ApJ...822...70L} support these assessments. The MDF of the MP stars in the outer halo peaks at a metallicity that is 0.5~dex lower than the MDF in the inner halo, while the peak of the MDF of the MR stars only varies by 0.1 dex between the two bins. We quantified the excess of MP stars with respect to the intermediate-metallicity (dotted blue line in Fig.~\ref{fig:density_final}) and the MR component (continuous red line in Fig.~\ref{fig:density_final}) with an exponential profile that is denoted by the dashed blue line in Fig.~\ref{fig:density_final}. The exponential profile has a central surface brightness of $\mu_{0} = 27.7\pm0.1\;\mathrm{mag}\;\mathrm{arcsec}^{-2}$ and a scale radius of $r_\mathrm{h} = 258 \pm 2 \arcsec$.

We next confirmed whether the exponential profile correctly predicts the fraction of MP stars ([M/H] $\leq -1.0$) in the inner \textit{HST} pointing, that is, at a major-axis distance of $\sim 250\arcsec$ from the centre of M105. We used the Gaussian MDFs from \citet[][Table~4 therein,]{2016ApJ...822...70L} and determined that $\sim18.6\%$ of the $2\,040$ stars in the MP component have metallicities [M/H] $\leq -1.0$. This corresponds to $\sim5\%$ of the $7\,445$ stars in the inner \textit{HST} pointing.
From the SB values of the exponential and total SB profiles defined above, $28.8\,\mathrm{mag}\,\mathrm{arcsec}^{-2}$ and  $25.7\,\mathrm{mag}\,\mathrm{arcsec}^{-2}$ , respectively, we expect $\sim5.8\%$ of the stars in the inner pointing to have metallicities [M/H] $\leq -1.0$. We again applied a metallicity-dependent conversion factor from SB to RGB counts derived from stellar population models based on the MILES stellar library (see Sect.~\ref{ssec:rgb2sb}). The exponential profile thus correctly predicts the fraction of MP stars in the inner halo.

With this information at hand, we now proceed to calculate the $\alpha$-parameter value associated with the MR and intermediate-metallicity RGB stars that dominate the number density profile in the inner halo. We assumed that the SB profile of this component is well modelled by the sum of the two S\'{e}rsic profiles with $n=2.8$ described in the previous paragraph. We determined the $\alpha$-parameter from the offset between the PN number density profile and the SB profile (using eq.~\eqref{eqn:density}, see Sect.~\ref{sec:pn_density} for details). This resulted in $\alpha_{2.5,\mathrm{MR+IM}} = (1.00\pm0.11) \times 10^{-8}\;\mathrm{PN}\;L^{-1}_\mathrm{bol}$. This value agrees well with the value determined for the inner halo solely based on surface photometry (cf. Sect.~\ref{sec:105_alpha}).

To determine the $\alpha$-parameter value associated with the MP RGB stars that dominate the number density profile in the outer halo, we constructed a two-component photometric model following the procedure outlined in Sect.~\ref{sec:M105photmodel}. In this model, the inner component is the sum of the two S\'{e}rsic profiles describing the MR and intermediate-metallicity RGB stars, and the outer component is the exponential profile determined earlier in this section. As in Sect.~\ref{sec:M105photmodel}, we did not fit the $\alpha$-parameter directly, but instead fit the ratio $\alpha_{2.5, \mathrm{MP}}/\alpha_{2.5,\mathrm{MR+IM}}$, corresponding to $\alpha_{2.5,\mathrm{MP}} = (7.10 \pm 1.87) \times 10^{-8}\;\mathrm{PN}\;L^{-1}_\mathrm{bol}$.

The thick black line in Fig.~\ref{fig:density_final} indicates the final best-fit model to the PN number density. The goodness-of-fit parameters are reported in Table~\ref{tab:fit_rgb}, and the posterior probability distribution of the model parameters is shown in Appendix~\ref{app:posterior}. The final model elegantly combines the information from integrated light \citep{2014ApJ...791...38W} resolved stellar populations \citep{2007ApJ...666..903H, 2016ApJ...822...70L} and PNe (this work) and provides a direct link between a high $\alpha$-parameter value and the emergence of a MP halo that follows an exponential profile.
The current finding for M105 is significant for the study of PNe in the galaxy halos and the surrounding IGL because it provides the unambiguous direct evidence that a large luminosity-specific PN number value is linked to the presence of a MP ([M/H] $< -1.0$) RGB population.

\section{Planetary nebula luminosity function of M105}
\label{sec:PNLFM105}
\begin{figure*}
  \centering
  \includegraphics[width = 8.79cm]{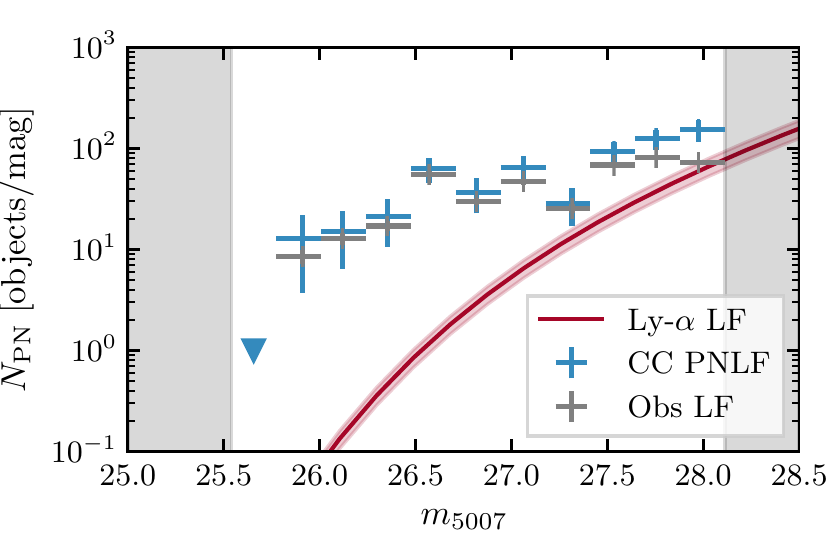}
  \includegraphics[width = 8.79cm]{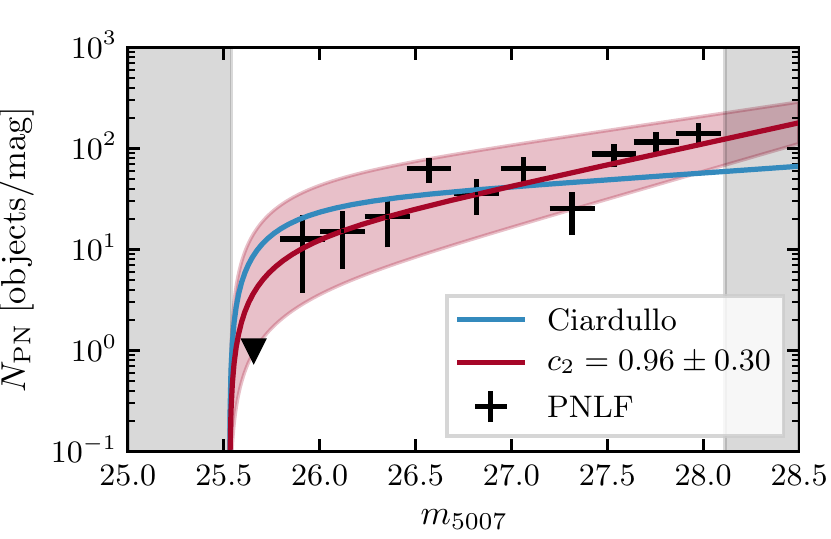}
  \caption{PNLF in M105. \textit{Left:} Observed LF of PN candidates (grey crosses) and PNLF of the completeness-corrected (CC) sample as defined in Fig.~\ref{fig:density_distr} (101 PNe) without accounting for contamination by Ly-$\alpha$ emitting background galaxies (blue crosses). Triangles denote upper limits. The solid red line shows the Ly-$\alpha$ LF and its variance due to density fluctuations indicated by the shaded regions \citep{2007ApJ...667...79G}. \textit{Right:} PNLF of the completeness-corrected PN sample after statistical subtraction of the Ly-$\alpha$ LF (black crosses). Triangles denote upper limits. The solid blue curve denotes the standard Ciardullo PNLF with $c_2 = 0.307$. The red line and shaded region denote our best-fit PNLF with $c_2 = 0.96 \pm 0.30$ In both panels, magnitudes fainter than the limiting magnitude and brighter than the bright cut-off are shaded in grey.}
  \label{figM105:pnlf_lyalpha}
\end{figure*}

In addition to the PN number density profile, the PNLF is an important characteristic of a PN sample and can be used to make inferences about the properties of the underlying stellar population. The bright cut-off of the PNLF is an important secondary distance indicator. In this section, however, we focus on the variation of the faint-end PNLF slope, as there has been empirical evidence for a variation of PNLF slope with stellar population properties \citep{2004ApJ...614..167C, 2010PASA...27..149C, 2013A&A...558A..42L, 2015A&A...579A.135L, 2017A&A...603A.104H, 2018A&A...616A.123H}. We therefore compared the observed PNLF to the generalised analytical formula introduced by \citet{2013A&A...558A..42L},
\begin{equation}
  N(M) = c_1 \mathrm{e}^{c_2 M} (1 - \mathrm{e}^{3 (M^{\star} - M)})
  \label{eq:pnlfgen}
.\end{equation}
In the equation above, $c_1$ is a normalisation factor and $c_2$ is the sought after faint-end slope. For $c_2 = 0.307,$ Eq. \eqref{eq:pnlfgen} takes the form of the standard PNLF introduced by \citet{1989ApJ...339...53C}.

We calculated the PNLF from the same sample that was used to determine the PN number density profile to limit contamination from PNe associated with NGC~3384 (see Fig.~\ref{fig:density_distr}).
The grey crosses in the left panel of Fig.~\ref{figM105:pnlf_lyalpha} denote the luminosity function (LF) of the PN candidates. The observed LF is contaminated by Ly-$\alpha$ emitting galaxies at redshift $z=3.1$ (see Appendix \ref{app:lyalpha}).
The estimated contribution of Ly-$\alpha$ emitters from \citet{2007ApJ...667...79G} is denoted by the red line and its expected 20\% variation by the red shaded region.
The observed number of PN candidates in our survey is larger than the estimated number of Ly-$\alpha$ contaminants in every magnitude bin.
After statistically subtracting the estimated number of contaminants, we applied the colour- and detection-completeness corrections (Table~\ref{tabM105:photometric}) and obtain the completeness-corrected (CC) PNLF shown with blue crosses in the left panel of Fig.~\ref{figM105:pnlf_lyalpha}. The crosses indicate the magnitude bin width and the expected errors due to counting statistics and the uncertainty of the Ly-$\alpha$ LF. For the brightest magnitude bin, the completeness-corrected number of PNe is smaller than one; we therefore only show an upper limit. The brightest PNe within 0.5 magnitudes from the bright cut-off of M105 are shown by red crosses in Fig.~\ref{fig:density_distr}.

In the right panel of Fig.~\ref{figM105:pnlf_lyalpha}, we again show the CC PNLF (black crosses) and different analytical PNLF models fit to the data.
A standard Ciardullo PNLF with a fixed $c_2 = 0.307$, as indicated by the solid blue line, is not a good representation of the data as it overestimates the number of bright PNe and underestimates the amount of faint PNe. We therefore fit the generalised PNLF as defined in Eq.~\eqref{eq:pnlfgen} to the observed PNLF. Because the distance to M105 is well determined from SBF measurements \citep[$D=10.3$~Mpc,][]{2001ApJ...546..681T}, the only free parameters are the normalisation $c_1$ and the faint-end slope, which we determine to be $c_2 = 0.96 \pm 0.30$. We used the non-linear least-squares minimisation method \textsc{lmfit}
 \citep{Newville_2014_11813} to fit to the completeness-corrected differential PNLF. The resulting analytic PNLF (denoted by the red line in the right panel of Fig.~\ref{figM105:pnlf_lyalpha}) reproduces the behaviour of the observed PNLF well. A PNLF with a steeper slope compared to M31 \citep{1989ApJ...339...53C} is clearly favoured.

\subsection{PNLF variation with radius}
\begin{figure*}
  \centering
  \includegraphics[width=8.79cm]{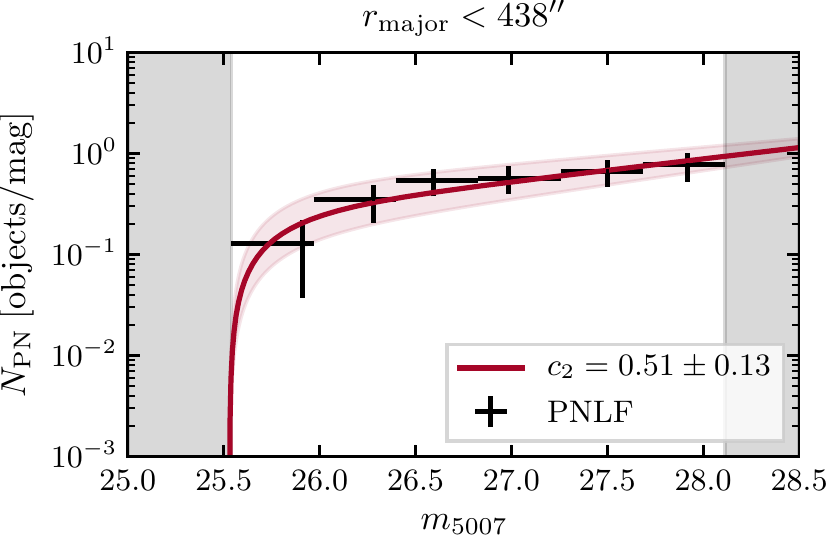}
  \includegraphics[width=8.79cm]{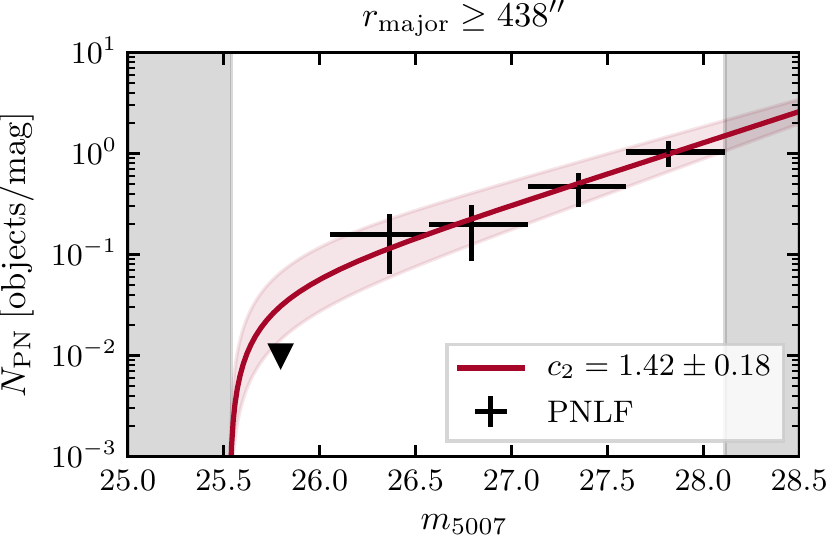}
  \caption{Radial variation of the PNLF in
  M105. \textit{Left}: Observed PNLF in the inner halo (black crosses) and
  best-fit generalised PNLF with $c_2 = 0.51\pm0.12$ (red line and shaded
  region). \textit{Right}: Observed PNLF in the outer halo (black crosses)
  and best-fit generalised PNLF with $c_2 = 1.42\pm0.18$ (red line and shaded
  region).}
  \label{fig105:PNLFradial}
\end{figure*}

Motivated by the evidence of the changing PN populations in the inner and outer halo based on the PN number density profile and the inferred $\alpha$-parameter values in Sect.~\ref{sec:105_alpha}, we investigated whether this difference also manifests itself in the PNLF morphology. We therefore divided our sample into two concentric elliptical bins and calculated the PNLF for each subsample individually. The dividing major-axis distance of $438\arcsec$ is indicated as the red ellipse in Fig.~\ref{fig:density_distr} and by the dashed vertical line in Fig.~\ref{fig:M105_density}.

The inner subsample\textup{} is characterised by a density profile that decreases steeply with radius and contains 63 PNe. Its PNLF is shown in the left panel of Fig.~\ref{fig105:PNLFradial}. A fit of the generalised PNLF (solid red line and shaded region) results in a value of $c_2 = 0.51\pm0.12$, which is shallower than the slope of the total PN sample. The outer subsample\emph{} is characterised by a flattened density profile and contains 38 PNe. Its PNLF is shown in the right panel of Fig.~\ref{fig105:PNLFradial}. A fit of the generalised PNLF (solid red line and shaded region) results in a slope of $c_2 = 1.42\pm0.18$, which is steeper than the slope of the total and inner PN samples.

The steeper PNLF in the outer subsample can be interpreted as an overabundance of faint PNe with respect to the inner subsample in the same magnitude bins, which may be the driving factor behind the flattening of the observed PN number density profile at large radii.
Based on the information from the resolved stellar populations discussed in Sect.~\ref{sect:RGB}, we can closely link the change in the PNLF morphology with the higher number of old and MP stars in the outer halo observed by \citet{2007ApJ...666..903H} and \citet{2016ApJ...822...70L}.
While there is strong evidence for a stellar population mix in the outer halo of M105, the PNLF beyond $r_\mathrm{major} \geq 438\arcsec$ does not deviate from a single generalised exponential LF with a cut-off at $M^{\star}$. \citet{2015a&a...575a...1r} fitted analytic two-mode PNLFs to PN populations in several nearby and Local Group galaxies, where a dip in the PNLF was observed that was several magnitudes fainter than the bright cut-off. \citet{2002aj....123..269j} argued that such dips could be interpreted as due to the existence of a population of PNe with faster central star evolution, for instance, for a younger stellar population. Similarly, \citet{2019A&A...624A.132B} observed an ubiquitous rise in the faint-end PNLF in M31, whose steepness can be inversely correlated with recent star formation \citep[][A\&A submitted]{SB2020}  A change in metallicity might have similar effects, but our data are not deep enough to explore them because PNLF dips have been observed to be at least two magnitudes fainter than the bright cut-off.
\section{Discussion and conclusions}
\label{sec:M105disc}
\subsection{Old, MP, and PN-rich outer halo of M105}

In Sect.~\ref{sec:105_alpha} we combined our study of individual tracers (PNe) with deep broad-band photometry from \citet{2014ApJ...791...38W} and found a strong flattening of the PN number density profile with respect to the stellar SB profile at large radii ($\gtrsim 9.8 r_\mathrm{e}$). This flattening can be well reproduced with a two-component photometric model that consists of a \emph{\textup{PN-poor}} inner halo with a de Vaucouleurs profile and a \emph{\textup{PN-rich}} exponential outer component. Relying solely on the information from surface photometry, we were unable to constrain the scale radius of the exponential profile.

We constrained the shape of the exponential profile using the information from resolved stellar populations. \citet{2007ApJ...666..903H} and \citet{2016ApJ...822...70L} published number-density profiles of MR and MP RGB stellar populations in the halo of M105. We combined the information from these studies to infer that the density distribution of the most MP stars ([M/H] $< -1.0$) follows an exponential profile with a central surface brightness of $\mu_{0} = 27.7\pm0.1\;\mathrm{mag}\;\mathrm{arcsec}^{-2}$ and a scale radius of $r_\mathrm{h} = 258 \pm 2 \arcsec$ (Sect~\ref{sect:RGB}).

Using the information from the RGB populations, we constructed a two-component photometric model, where the two components were weighted by their $\alpha$-parameter values. In the  \emph{\textup{PN-poor}} inner halo, the light distribution is dominated by intermediate-metallicity ($-1.0 \le \mathrm{[M/H]}< -0.5)$ and MR  ([M/H] $\ge -0.5]$) stars, while in the \emph{\textup{PN-rich}} outer halo the MP stars dominate the light distribution. The $\alpha$-parameter values of the two components are as follows:
\begin{itemize}
  \item $\alpha_\mathrm{2.5,MR+IM} = (1.00 \pm 0.11)\times 10^{-8}\;\mathrm{PN}\;L^{-1}_\mathrm{bol}$ for the inner \emph{\textup{PN-poor}} component, and
  \item $\alpha_\mathrm{2.5,MP} = (7.10 \pm 1.87)\times 10^{-8}\;\mathrm{PN}\;L^{-1}_\mathrm{bol}$ for the outer \emph{\textup{PN-rich}} component.
\end{itemize}
We find a difference of a factor of seven in the $\alpha$-parameter values of the inner versus outer components, implying that the luminosity-specific PN number of the stellar population that traces the old and MP halo at large radii is seven times higher.

We interpret the high $\alpha$-parameter value and the steep faint-end PNLF slope as tell-tale signs of an additional old and MP stellar population whose contribution towards the overall light distribution increases at large radii. This interpretation is corroborated by the results of \citet{2007ApJ...666..903H} and \citet{2016ApJ...822...70L} based on resolved stellar populations.
While in earlier works \citep{2013A&A...558A..42L, 2017A&A...603A.104H} this association was made mostly by inference, we are now able to \emph{\textup{directly}} establish the link between the high $\alpha$-parameter value, the steep PNLF slope, and the increase in the old and MP stellar population in the outer regions of M105 \citep[identified by][]{2007ApJ...666..903H}.

The data of \citet{2007ApJ...666..903H} clearly ruled out a single chemical evolution sequence for the entire halo stellar population and instead suggested a multi-stage evolutionary model. For the outer MP halo, two different scenarios were proposed: \citet{2007ApJ...666..903H} argued that the MP halo was built up from pristine gas and that star-formation therein was truncated early, at the time of cosmological reionisation. In contrast, \citet{2016ApJ...822...70L} proposed a formation dominated by dissipationless mergers and accretion. The scenario by \citet{2007ApJ...666..903H} implies that the faint MP halo reached the current extension out to 40-50 kpc at high redshift, which is at odds with the observed slow growth of the stellar half-mass radius as a function of redshift \citep{2008ApJ...688...48V}. The scenario proposed by \citet{2016ApJ...822...70L} is more consistent with the current two-phase formation model for elliptical galaxies.

Based on our current photometric PN survey in M105, we cannot yet fully distinguish between these two scenarios. The high $\alpha$-parameter value associated with the old and MP population is similar to the $\alpha$-parameter values of Local Group dwarf irregular galaxies such as Leo~I, Sextans A, and Sextans B, which lie between $3\times10^{-8}$ and $10^{-7}\;\mathrm{PN}\;L^{-1}_\mathrm{bol}$ \citep{2006MNRAS.368..877B} and are in line with the accretion scenario proposed by \citet{2016ApJ...822...70L}. The new kinematics from the PN.S may provide viable information because we would be able to measure a higher velocity dispersion in the outer halo if it were accreted. Recent results on the outer halos of M49 and M87 \citep{2018A&A...616A.123H, 2018A&A...620A.111L} definitely support the two-phase formation scenario.

\subsection{Extended halo or IGL of the Leo~I group?}

The results from our PN photometry point towards an extended but rather faint outer stellar envelope that dominates the observed PN number density profile at large radii from the centre of M105, which is one of the group-dominant galaxies in the Leo~I group. The SB of the exponential outer component that we associate with these PNe falls below the detection limit of $\mu_B = 30 \; \mathrm{mag}\;\mathrm{arcsec}^{-2}$ of \citet{2014ApJ...791...38W} beyond a distance of 750\arcsec. It is therefore not surprising that \citet{2014ApJ...791...38W} concluded that there was no outer component in addition to the halo of M105.

Based on the vastly different $\alpha$-parameter value associated with the MP stellar populations and the distinct PNLF slope, we can already confirm a distinct origin of the stellar population associated with this outer exponential profile. Further independent evidence in support to the IGL classification of this component comes from the colour-magnitude diagram (CMD) of the resolved RGB stars in the western HST field \citep{2007ApJ...666..903H,    2016ApJ...822...70L}: it has the same properties and MDF as the CMD and the corresponding MDF of the resolved intra-cluster RGB stars tracing the ICL in the Virgo core \citep{2007ApJ...656..756W}.

We can further identify common traits between the PN population and integrated-light properties at large radii in M105 compared to other group- and cluster-dominant galaxies. As we discussed in the previous paragraphs, the strong increase in the $\alpha$-parameter value at large radii and the exponential SB profile for the PN-rich component were also observed for the ICL/IGL surrounding cluster and group-dominant galaxies. For example, the exponential profile around M105 has a quite similar scale length $r_\mathrm{h}$ to that in M49, see \citet{2017A&A...603A..38S}, while its central SB $\mu_0$ value is more than a magnitude fainter in M105 than M49. When the values of the structural parameters for the exponential component in M105 are compared with those of the outer exponential components identified in more massive systems such as the BCGs NGC~3311 in the Hydra cluster \citep{2012A&A...545A..37A} or NGC~1399 in the Fornax cluster \citep{2016ApJ...820...42I}, the $\mu_0$ values increases by several magnitudes and the scale radius decreases by a factor two, signalling an increase in total luminosity and hence light fraction of the diffuse outer component with mass.

Returning to M105, we can estimate the IGL fraction, assuming that the IGL is traced by the PNe and RGB stars associated with the exponential SB profile that we fit to the MP stars. We first estimated the bolometric luminosity of the diffuse exponential component. When we integrate the exponential profile over the total area of our SuprimeCam survey, we obtain $L_\mathrm{bol,MP} = 2.04 \times 10^9\;L_{\odot}$. Within the limiting magnitude of our survey, we would hence expect $145$ PNe to be associated with the exponential outer component. For comparison, the total bolometric luminosity associated with the main halo within our survey of M105 is $L_\mathrm{bol,MR+IM} = 5.1\times10^{10}\;L_{\odot}$.
This luminosity would correspond to over $510$ PNe, assuming an $\alpha$-parameter value equal to $\alpha_\mathrm{2.5,MR+IM}$ as stated above. Because of incompleteness, especially in the bright galaxy centre, we observed significantly fewer PNe.
We therefore determine the fraction of PNe associated with the outer component to be $22\%$. We note that this number is higher than that determined by  \citet{2003A&A...405..803C} in the vicinity of the \ion{H}{i} ring.
The fraction of the bolometric luminosities of IGL versus the total light in galaxies is thus $3.8\%$, which is larger than the upper limit stated by \citet{2014ApJ...791...38W}. We note that the luminosity estimate of the diffuse component compares well with that of single ultra-diffuse galaxies (UDGs) observed in groups and clusters of galaxies \citep{2015ApJ...809L..21M}. UDGs are one viable progenitor of IGL stars: as a result of their low mass and density, they can be easily stripped, and their material can be deposited at large radii in the halo of a massive ETG \citep{2017MNRAS.464.2882A}.

In Paper~II of this series we will address the origin of the IGL by evaluating the dynamical status of the PNe populations at large radii based on the recently obtained kinematics in the framework of e$^{2}$PN.S ETG survey. This will then also enable us to draw comparisons between the IGL/ICL kinematics in different environments and compare these results with recent integral-field spectroscopic surveys of massive ETGs \citep[e.g.][]{2014ApJ...795..158M}. The assessment of the IGL in the Leo~I group is particularly relevant because this group represents a benchmark for the IGL properties at the low-mass end of the galaxy group and cluster mass function.

\subsection{Outlook}
We have presented evidence that PNe trace a faint, MP component around M105 in the Leo~I group on the basis of deep photometry and number-count profiles of PNe and resolved stellar populations.
\citet{2019arXiv190808544K} surveyed hundreds of brightest cluster galaxies (BCGs) at low redshift ($z < 0.08$) and fitted their SB profiles with double S\'{e}rsic models. They found a wide scatter between the transition radius of the two S\'{e}rsic profiles and the corresponding SB at this radius, which led them to the conclusion that the outer S\'{e}rsic profile probably does not trace the ICL. Even though we used a different functional decomposition, the conclusion of \citeauthor{2019arXiv190808544K} might be applied to M105 and its exponential outer component.
Analyses of numerical simulations have shown that it is challenging to distinguish IGL or ICL components from outer envelopes that are gravitationally bound to the galaxy based on SB profiles alone \citep[e.g.][]{2015MNRAS.451.2703C}.
In order to associate the diffuse component surrounding M105 with the IGL, we need to know whether it is gravitationally bound to the group potential. This question cannot be answered based on photometric surveys alone.

In Paper~II of this series, the analysis of the PN kinematics based on a photometric and kinematic decomposition of M105 and NGC~3384 will be presented. These new data will provide first constraints on the physical state of the extended exponential outer component and the surrounding IGL identified here.

\section*{Acknowledgements}
The authors thank S. Okamura for his contribution towards the Subaru observations.
We greatly acknowledge the support and advice of the ING and Subaru staff. We also thank A.~E. Watkins and J.~C. Mihos for the provision of the tabulated surface-brightness and colour profiles of M105 and NGC~3384 and M.~G. Lee and I.~S. Jang for the provision of the density profiles of the resolved RGB stellar populations in the halo of M105. We thank the organisers of the ``Light in the suburbs'' conference for the many stimulating discussions during the conference.
JH thanks A. Agnello, I. S\"{o}ldner-Rembold, and C. Spiniello for useful comments and discussions. JH and CP furthermore acknowledge support from the IMPRS on Astrophysics at the LMU Munich. We thank the referee Marc Sarzi for his constructive comments and the careful reading of the manuscript.

This research made use of \textsc{astropy} \citep{2013A&A...558A..33A},
\textsc{astroplan} \citep{2018AJ....155..128M},
\textsc{astroquery} \citep{2019arXiv190104520G},
\textsc{astromatic-wrapper},
\textsc{corner} \citep{ForemanMackey2016},
\textsc{emcee} \citep{2013PASP..125..306F},
\textsc{lmfit} \citep{Newville_2014_11813},
\textsc{matplotlib} \citep{Hunter:2007},
\textsc{numpy} \citep{2011arXiv1102.1523V},
and \textsc{sympy} \citep{10.7717/peerj-cs.103}.
This research has made use of the SIMBAD database, operated at CDS, Strasbourg, France \citep{2000A&AS..143....9W}. This research has made use of the VizieR catalogue access tool, CDS,
Strasbourg, France.  The original description of the VizieR service was published in \citet{2000A&AS..143...23O}. This publication has made use of data products from the Two Micron All Sky Survey, which is a joint project of the University of Massachusetts and the Infrared Processing and Analysis Center/California Institute of Technology, funded by the National Aeronautics and Space Administration and the National Science Foundation. The Digitized Sky Surveys were produced at the Space Telescope Science Institute under U.S. Government grant NAG W-2166. The images of these surveys are based on photographic data obtained using the Oschin Schmidt Telescope on Palomar Mountain and the UK Schmidt Telescope. The plates were processed into the present compressed digital form with the permission of these institutions.
This research has made use of NASA's Astrophysics Data System.
\bibliographystyle{aa} 
\bibliography{literature}

\appendix
\section{Extraction of the photometric PN catalogue}
\label{app:catalog_extraction}
We used Source Extractor \citep[SExtractor,][]{1996A&AS..117..393B} to detect objects, measure their photometric properties, and to calculate background maps of our mosaic on- and off-band images. We denote the measured narrow- and broad-band magnitudes with $m_\mathrm{n}$ and $m_\mathrm{b}$ , respectively. The extracted magnitudes are in the $AB$ magnitude system.

SExtractor detected and analysed objects for which the flux value in at least nine adjacent pixels was higher than $1.1\times\sigma_\mathrm{rms}$ on the on-band image. As we aimed to measure the colour $m_\mathrm{n} - m_\mathrm{b}$ of each emission-line object, we chose an aperture radius of 13~pixels and subsequently extracted the broad-band magnitude $m_\mathrm{b}$ on the off-band image in dual-image mode, that is, in the same aperture as in the on-band image. Following \citet{1997MNRAS.284L..11T}, we assigned a magnitude of $m_\mathrm{b} = 28.7$ to [\ion{O}{iii}] emitters for which no $V$-band magnitude could be measured in the off-band image.
This corresponds to the flux of an [\ion{O}{iii}] emitter at the limiting magnitude $m_\mathrm{n,lim}$ observed through a broad-band $V$ filter.

\subsection{Catalogue pre-processing}
\label{app:preprocess}
To prevent false detections of PNe due to image artefacts, we masked the pixel columns affected by dithering as well as regions with a high background value (e.g. due to charge transfer or saturated stars). We calculated the rms-background maps of the on- and off-band images using SExtractor and masked all pixels with values higher than 1.5~times the median background. Furthermore, we masked the central regions of the three bright galaxies M105, NGC~3384, and NGC~3398 due to high background counts from diffuse galaxy light (black ellipses in Fig.~\ref{fig:ly-test-distr}). The masked regions account for $2.3\%$ of the total survey area, which results in an unmasked survey area that covers $0.2365\,\mathrm{deg}^2$.

\subsection{Limiting magnitude and photometric error}
\label{app:limmag}
We determined the limiting magnitude of our survey by populating the on-band image with a synthetic PN population and determining its recovery fraction as a function of magnitude. The simulated PNe have a Moffat-shaped PSF profile as measured on the image (see Sect.~\ref{ssec:photsurvey}) and follow the standard PNLF \citep{1989ApJ...339...53C} scaled to the distance of M105. The synthetic population was added to the on-band image using the IRAF task \texttt{mkobjects}. The limiting magnitude $m_\mathrm{lim}$ is defined as the magnitude at which the recovery fraction of the simulated population falls below $50\%$. We determine a limiting magnitude of $m_{\mathrm{lim},n} = 25.6$, which corresponds to $m_{\mathrm{lim},5007} = 28.1$.

We also used the synthetic population to determine the error on the extracted magnitudes by determining the difference between the simulated magnitudes and those extracted using SExtractor. We estimate the photometric error to be one standard deviation about the mean in ten magnitude bins. The resulting error bands are shown in Fig.~\ref{figM105:mag-mag}.

\subsection{Object selection}
\label{ssec:objsel}
PN candidates were selected based on their position on the CMD and their unresolved light distribution.

\subsubsection{CMD selection}
\begin{figure}
  \centering
    \includegraphics[width=8.8cm]{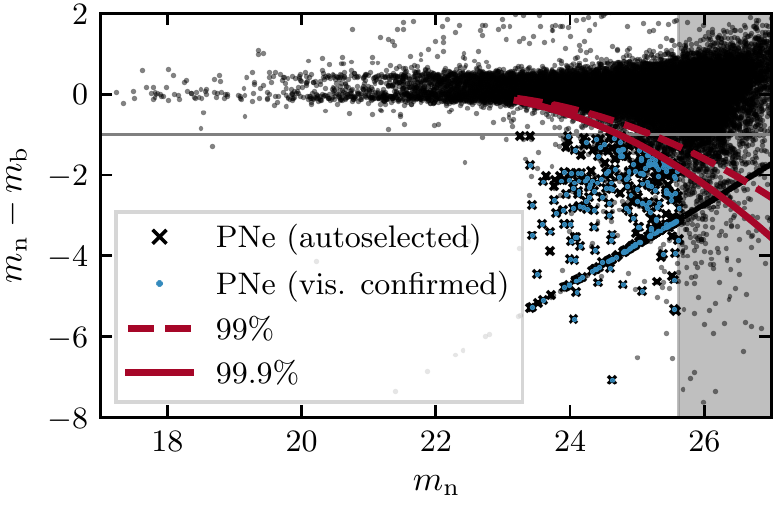}
  \caption{CMD for all sources (black dots) in the M105 Suprime Cam field. Obects that lay below the grey horizontal line at $m_\mathrm{n} - m_\mathrm{b} = -1$ are emission-line objects with an equivalent width $\mathrm{EW}_\mathrm{obs} \geq 110\,\angstrom$. The magnitude range that is fainter than the limiting magnitude is denoted by the shaded grey region. The dashed (solid) red lines denote the $99.9\%$ ($99.9\%$) limits of the simulated foreground population. Black crosses denote the PNe candidates that were automatically selected, and blue dots denote those that were confirmed by visual inspection.}
  \label{figM105:cmd}
\end{figure}

Figure~\ref{figM105:cmd} shows the $m_\mathrm{n}$ versus $m_\mathrm{n} - m_\mathrm{b}$ CMD of all extracted sources in the M105 Suprime Cam field. In order to limit the contamination by [\ion{O}{ii}]$3727\,\angstrom$ emitters at redshift $z = 0.345$, we imposed a colour-selection criterion of $m_\mathrm{n} - m_\mathrm{b} < -1$. This corresponds to an observed equivalent width of $EW_\mathrm{obs} > 110\,\angstrom$ \citep{2000ApJ...542...18T}. No [\ion{O}{ii}] emitters with $EW_\mathrm{obs} > 110\,\angstrom$ have been observed to date \citep{1990MNRAS.244..408C,1997ApJ...481...49H,1998ApJ...504..622H}.
PNe candidates are thus objects that have a colour excess $m_\mathrm{n} - m_\mathrm{b}$ and are brighter than the limiting magnitude $m_{\mathrm{lim},n} = 25.6$.

Another source of possible contaminants are faint stellar sources (e.g. MW stars) that are scattered below the adopted colour excess because the increasingly large photometric errors are faint magnitudes. We therefore simulated a continuum population onto the on- and off-band images. The red dashed and solid lines in Fig.~\ref{figM105:cmd} denote the $99\%$ and $99.9\%$ limits calculated from the extracted CMD of the simulated population. To limit the probability of including foreground stars to the $0.1\%$ level, we only considered objects below the $99.9\%$ limit for further analysis.

\subsubsection{Point-like versus extended sources}
\begin{figure}
  \centering
    \includegraphics[width=8.8cm]{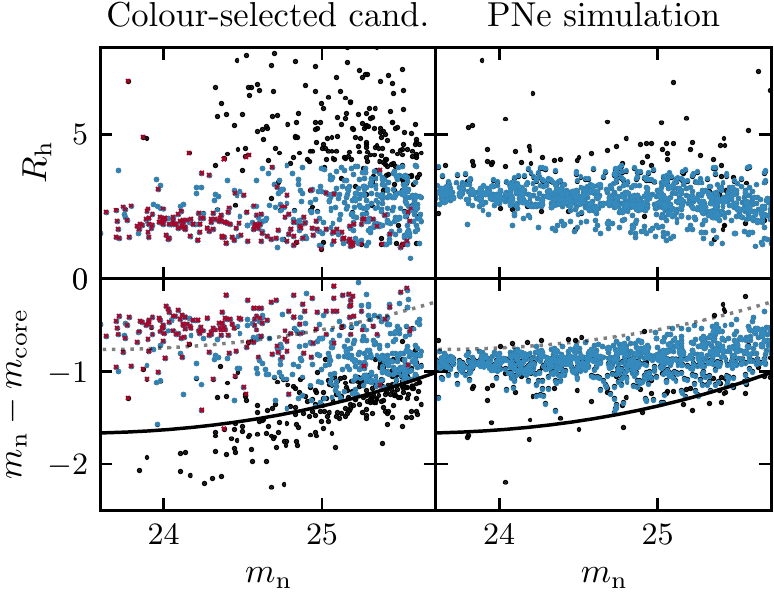}
  \caption{Point-source test for PN candidates in M105. \emph{Left column:} Colour-selected PNe candidates. \emph{Right column:} Simulated PNe. \emph{Top row:} Half-light radius $R_\mathrm{h}$ as function of magnitude $m_\mathrm{n}$. \emph{Bottom row:} $m_\mathrm{n} - m_\mathrm{core}$ as function of magnitude $m_\mathrm{n}$. In each panel, blue dots denote the PNe candidates selected based on their point-like appearance. In the left column, red crosses denote cross-matches with the PN.S catalogue, which place additional constraints on the morphology of the selected PNe candidates from Subaru photometry.}
  \label{figM105:pointlike}
\end{figure}

At extragalactic distances, PNe appear as unresolved point-like sources. To limit contamination from extended sources (e.g. background galaxies or other objects with strong [\ion{O}{iii}] emission), we analysed the light distribution of the simulated PN population. We derived the following criteria for point-like sources:
\begin{enumerate}
  \item The sources have half-light radii in the range $1 < R_\mathrm{h} < 3.9\,\mathrm{pix}$ (top panels of Fig.~\ref{figM105:pointlike}). This upper limit corresponds to $95\%$ of the simulated population.
  \item The sources lie above the $95\%$ limit derived from the $m_\mathrm{n}$ versus $m_\mathrm{n} - m_\mathrm{core}$ distribution of the simulated population (solid black line in the bottom panels of Fig.~\ref{figM105:pointlike}).
  The core magnitude $m_\mathrm{core}$ is measured within a $5\,\mathrm{pix}$ aperture.
\end{enumerate}
Sources that fulfil both criteria are denoted in blue in Fig.~\ref{figM105:pointlike} and as black crosses on the CMD in Fig.~\ref{figM105:cmd}.
In contrast to \citet{2013A&A...558A..42L} and \citet{2017A&A...603A.104H}, we lifted the upper limit on the $m_\mathrm{n}$ versus $m_\mathrm{n} - m_\mathrm{core}$ distribution (grey dashed line in Fig.~\ref{figM105:pointlike} ). This was motivated by the distribution of the PN.S-matched sources (red crosses) in the bottom left panel of Fig.~\ref{figM105:pointlike}, the majority of which lie above the dashed grey line.

\subsection{Catalogue validation and visual inspection}
The automatically selected photometric catalogue was visually inspected, and any remaining spurious detections were removed, resulting in a final catalogue of 226 sources within the limiting magnitude of $m_{\mathrm{lim},n} = 25.6$. These sources are highlighted with blue dots in the CMD in Fig.~\ref{figM105:cmd}. Their spatial distribution is shown in Fig.~\ref{figM105:survey}, superposed on the DSS image, in which they are denoted by blue crosses.

\subsection{Catalogue completeness}
\label{app:compl}

Because PN-candidates were selected based on flux-ratio criteria, PNe that
fell below the flux-threshold due to photometric errors would not have been
detected. Table~\ref{tabM105:spatial} shows the \emph{\textup{spatial completeness}} of
our survey within its limiting magnitude as a function of major-axis radius. The
spatial completeness was calculated by determining the recovery fraction of the
synthetic PN population simulated onto the unmasked regions of the on-band
image. The spatial completeness is lowest in the innermost bin, which is due to
the high background counts from the diffuse galaxy light of M105.

The \emph{\textup{photometric completeness}} as a function of magnitude can be evaluated
in terms of colour and detection completeness. The latter is simply the
recovery fraction of the simulated population from the unmasked regions of the
on-band image as a function of magnitude (cf. Appendix.~\ref{app:limmag}). We
determined the colour completeness by evaluating how many simulated PNe were
scattered to CMD regions outside of the colour-selection criteria because of
photometric errors as a function of magnitude. Both colour and detection
completeness are tabulated in  Table~\ref{tabM105:photometric}.

\subsection{Possible sources of contamination}
\subsubsection{Faint continuum objects}
\label{app:faint}

As we described earlier (cf. Appendix.~\ref{ssec:objsel}), we limited our
selection of objects to those that fall below the $99.9\%$ line of a simulated
continuum population to minimise the contamination by foreground
objects such as faint MW stars. However, some of these objects will be scattered
below this line. We estimated the fraction of foreground contaminants in the
colour-selected sample by determining the total number of observed foreground
stars down to $m_\mathrm{n,lim}$ and assuming that $0.1\%$ of these are
scattered into the sample of colour-selected PNe. This results in $23$ objects,
which corresponds to a contribution of $10\%$ of the extracted sample of 226
PNe.

\subsubsection{Ly-$\alpha$ emitting background galaxies}
\label{app:lyalpha}
Another source of contamination are faint background Ly-$\alpha$ galaxies that emit at the same wavelength as PNe in the Leo~I group, if at redshift $z = 3.1$. To quantify this effect, we used the Ly-$\alpha$ LF determined by \citet{2007ApJ...667...79G}, who carried out a deep survey for $z = 3.1$ Ly-$\alpha$ emission-line galaxies. The LF is characterised by a Schechter function \citep{1976ApJ...203..297S}. More recent surveys \citep[e.g.][]{2012ApJ...744..110C} agree with this LF within $0.1$ mag. The large-scale clustering of the Ly-$\alpha$ population at redshift $z=3.1$ has a correlation length of $r_0 = 3.6$~Mpc \citep{2007ApJ...671..278G}. At the distance of M105, this corresponds to an angular scale of $20\fdg16$, which is much larger than the SuprimeCam FoV.
Within the limiting magnitude of our survey, we expect $56$ Ly-$\alpha$ emitters, which corresponds to  $24.9\pm5.0\%$ of the completeness-corrected sample. Figure~\ref{figM105:pnlf_lyalpha} shows the corresponding LF.
The contribution from [\ion{O}{ii}]$3727\angstrom$ emitters at redshift $z = 0.345$ is already accounted for when the \citet{2007ApJ...667...79G} Ly-$\alpha$ LF is used \citep[cf.][]{2013A&A...558A..42L, 2017A&A...603A.104H}.

\begin{table}
  \centering
\caption{Spatial completeness and unmasked survey area as function of major-axis radius.}
\begin{tabular}{lll}
\hline
\hline
$r_\mathrm{major}$ & $A$ & $c_\mathrm{spatial}$ \\
$[\mathrm{\degr}]$ & $[\mathrm{\degr}^2]$ &  \\
\hline\T
0.0454 & 0.0013 & 0.6471 \\
0.0603 & 0.0023 & 1.0000 \\
0.0705 & 0.0022 & 1.0000 \\
0.1102 & 0.0114 & 0.9683 \\
0.1366 & 0.0098 & 0.7963 \\
0.1819 & 0.0228 & 0.7869 \\
0.2175 & 0.0230 & 0.7583 \\
0.3554 & 0.0684 & 0.8649 \\
\hline
\end{tabular}
\label{tabM105:spatial}
\end{table}

\begin{table}
  \centering
\caption{Colour and detection completeness as function of magnitude.}
\begin{tabular}{lll}
\hline
\hline
$m_\mathrm{n}$ & $c_\mathrm{colour}$ & $c_\mathrm{detection}$ \\
$[\mathrm{mag}$] &  &  \\
\hline\T
23.265 & 0.66 & 0.64 \\
23.565 & 0.94 & 1.00 \\
23.738 & 0.94 & 0.72 \\
23.897 & 0.87 & 0.96 \\
24.031 & 0.89 & 1.00 \\
24.146 & 0.87 & 1.00 \\
24.260 & 0.91 & 0.83 \\
24.376 & 0.89 & 0.96 \\
24.476 & 0.87 & 0.96 \\
24.594 & 0.96 & 0.72 \\
24.714 & 0.83 & 0.96 \\
24.808 & 0.91 & 1.00 \\
24.894 & 0.89 & 0.96 \\
24.989 & 0.85 & 0.87 \\
25.087 & 0.87 & 0.83 \\
25.182 & 0.81 & 0.81 \\
25.277 & 0.76 & 0.86 \\
25.363 & 0.85 & 1.00 \\
25.450 & 0.72 & 0.74 \\
25.561 & 0.70 & 0.56 \\
\hline
\end{tabular}
\label{tabM105:photometric}
\end{table}

\section{Posterior probability distributions of the two-component model parameters}
\label{app:posterior}
\begin{figure}
  \includegraphics[width = 8.8cm]{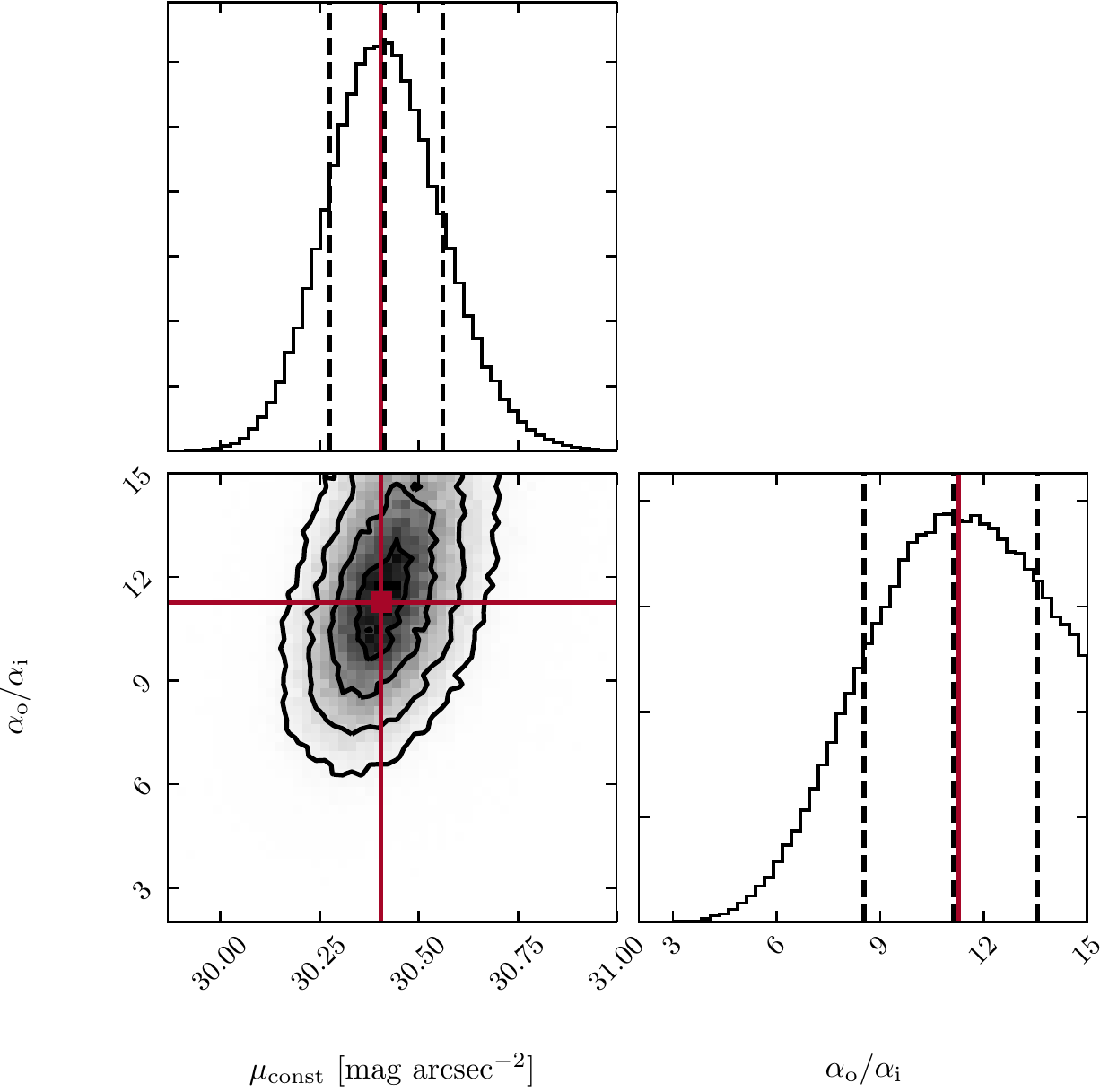}
  \caption{Posterior probability distribution of the parameters $\mu_\mathrm{const}$ and $\alpha_\mathrm{outer}/\alpha_\mathrm{inner}$ of the two-component model with constant outer SB profile. The red lines denote the best-fit parameters determined in Sect.~\ref{sec:M105photmodel} (see Table~\ref{tab:fit_res}), and the dashed lines show the $16\%$, $50\%$, and $84\%$ levels.}
  \label{fig:const_corner}
\end{figure}

\begin{figure}
  \includegraphics[width = 8.8cm]{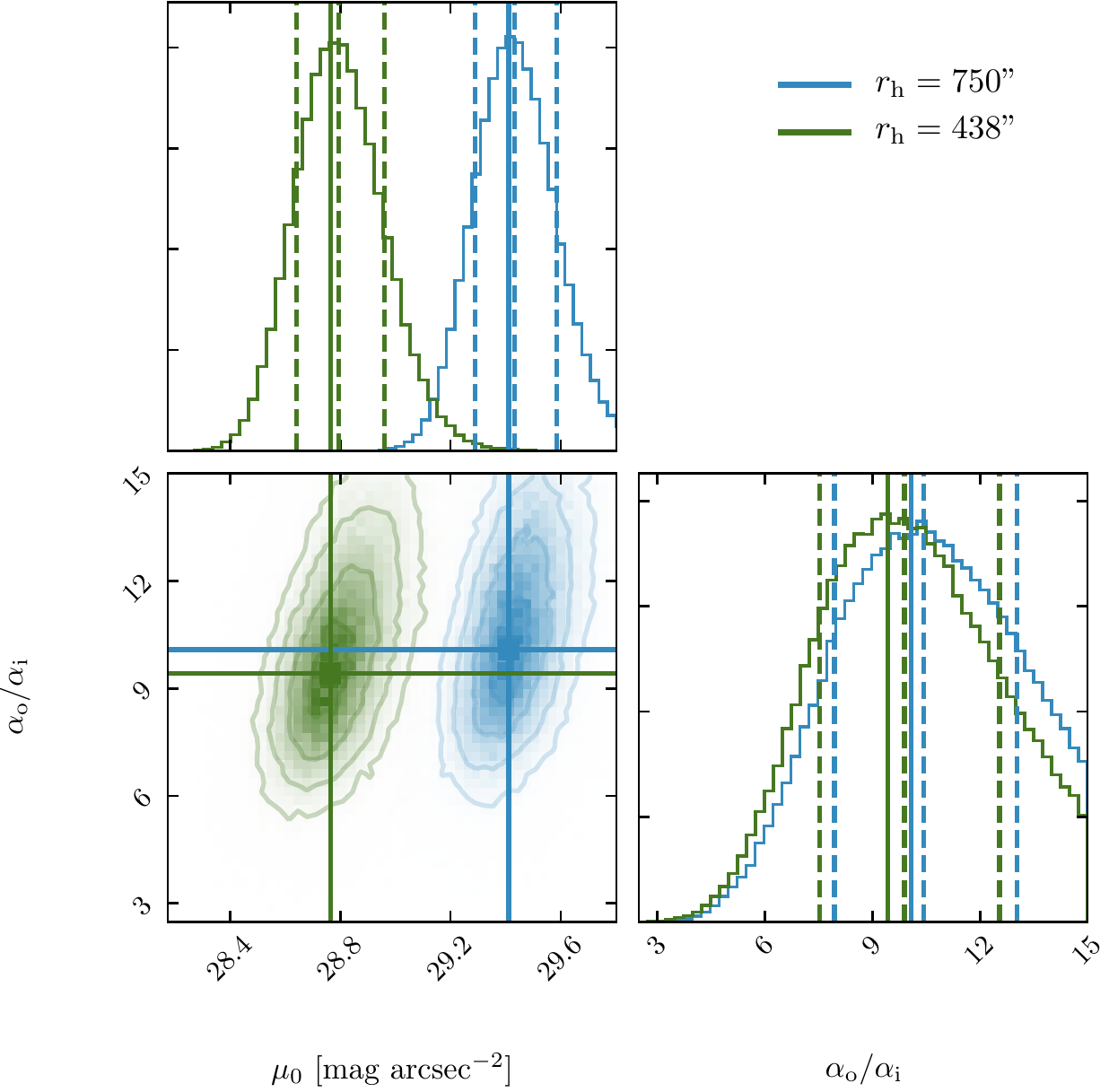}
  \caption{Posterior probability distribution of the parameters $\mu_\mathrm{const}$ and $\alpha_\mathrm{outer}/\alpha_\mathrm{inner}$ of the two-component model with an exponential outer SB profile. The distribution shown in blue corresponds to a model with a fixed $r_\mathrm{h} = 750\arcsec$ , and the distribution shown in green corresponds to a fixed $r_\mathrm{h} = 438\arcsec$. The solid lines denote the best-fit parameters determined in Sect.~\ref{sec:M105photmodel} (see Table~\ref{tab:fit_res}), and the dashed lines show the $16\%$, $50\%$, and $84\%$ levels.}
  \label{fig:exp_corner}
\end{figure}

\begin{figure}
  \includegraphics[width = 8.8cm]{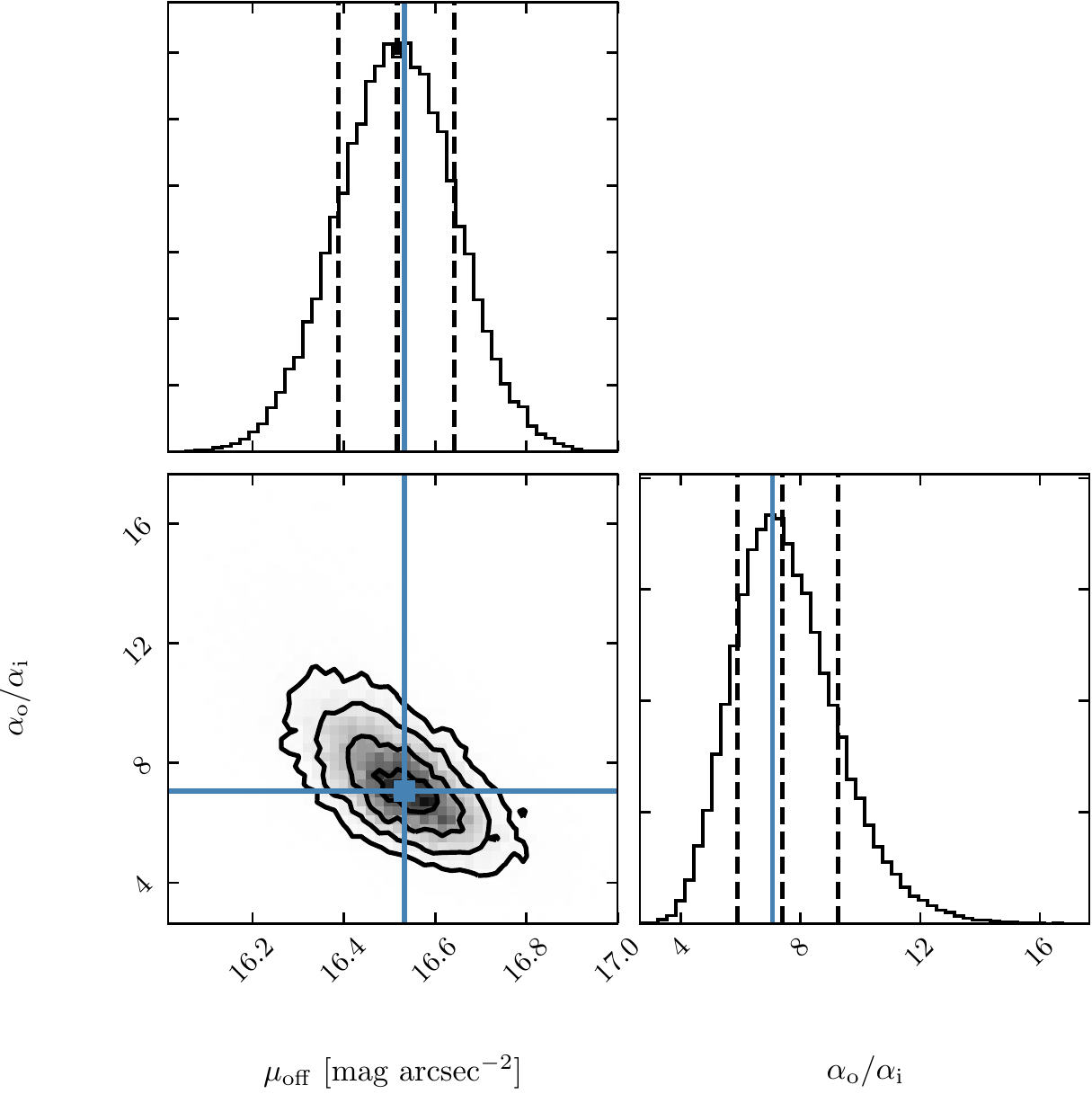}
  \caption{Posterior probability distribution of the parameters $\mu_\mathrm{off}$ and $\alpha_\mathrm{outer}/\alpha_\mathrm{inner}$ of the two-component model with an exponential outer SB profile fit determined from the RGB number density profiles. The blue lines denote the best-fit parameters determined in Sect.~\ref{sect:RGB} (see Table~\ref{tab:fit_rgb}), and the dashed lines show the $16\%$, $50\%$, and $84\%$ levels.}
  \label{fig:exp_corner_final}
\end{figure}

We calculated posterior probability distributions of the model parameters for each of the models presented in Sect.~\ref{sec:M105photmodel}. We used the following logarithmic likelihood function \citep{2013PASP..125..306F}:
\begin{equation}
\ln p(D|F_\mathrm{true}) = -\frac{1}{2}\sum_n \left[\frac{(g_n(F_\mathrm{true}) - D_n)^2}{s_n^2}+\ln (2\pi s_n^2)\right],
\end{equation}
where $D$ is the data, that is, the observed stellar SB and PN number density profiles, $F$ is the model parameters as detailed in Table~\ref{tab:fit_res}, $g$ is the model function (Eqs.~\eqref{eq:M105stellarmodel} and \eqref{eq:M105photmodel}), and $s_n$ is the measurement uncertainties of the data.
The resulting posterior probability distributions, calculated with the ensemble-based Monte Carlo Markov chain (MCMC) sampler \textsc{emcee} (Foreman-Mackey et al. 2013), are shown in Fig.~\ref{fig:const_corner} for the model with the constant outer component, and Fig.~\ref{fig:exp_corner} shows the two models with an exponential outer component. Figure~\ref{fig:exp_corner_final} shows the posterior probability distribution for the final best-fit model with priors from resolved stellar populations. For each model, the posterior probability distribution was evaluated using $100$ walkers and $5\,000$ steps.

\end{document}